\documentclass[nologo,url,11pt,a4paper]{ETHpaper}
\usepackage{amssymb,amsmath} 
\usepackage{subfig}
\usepackage[sort&compress]{natbib} 
\usepackage[utf8]{inputenc}
\usepackage{graphicx}
\usepackage{pict2e}
\usepackage{bbm}
\usepackage{todonotes}
\usepackage[export]{adjustbox}

\begin{document}

\title{Systemic risk in multiplex networks \\ with asymmetric coupling and threshold feedback}
\titlealternative{Systemic risk in multiplex networks with asymmetric coupling and threshold feedback}
\author{Rebekka Burkholz, Matt V. Leduc, Antonios Garas \& Frank Schweitzer}
\authoralternative{R. Burkholz, M.V. Leduc, A. Garas \& F. Schweitzer}
\address{Chair of Systems Design, ETH Zurich, Weinbergstrasse 58, 8092
   Zurich, Switzerland}

\reference{}

\www{\url{http://www.sg.ethz.ch}}

\makeframing
\maketitle

\begin{abstract}
We study cascades on a two-layer multiplex network, with asymmetric feedback that depends on the coupling strength between the layers.  
Based on an analytical branching process approximation, we calculate the systemic risk measured by the final fraction of failed nodes on a reference layer. 
The results are compared with the case of a single layer network that is an aggregated representation of the two layers. 
We find that systemic risk in the two-layer network is smaller than in the aggregated one only if the coupling strength between the two layers is small. 
Above a critical coupling strength, systemic risk is increased because of the mutual amplification of cascades in the two layers. 
We even observe sharp phase transitions in the cascade size that are less pronounced on the aggregated layer. 
Our insights can be applied to a scenario where firms decide whether they want to split their business into a less risky core business and a more risky subsidiary business. In most cases, this may lead to a drastic increase of systemic risk, which is underestimated in an aggregated approach. 
\end{abstract}

\section{Introduction}
\label{sec:intro}

Cascading failures in complex systems can be understood as a process by which the initial failure of a small set of individual components leads to the failure a significant fraction of the system's components. This is due to interconnections between the different components of the system. Such a phenomenon can occur in physical systems such as power grids (e.g. \cite{Carreras2004}, \cite{Kinney2005} and \cite{Brummitt2012}), but also in complex organizations like interbank systems (e.g. \cite{EisenbergNoe2001}, \cite{Gai2010}, \cite{Battiston2012a} and \cite{Amini2010}). A general framework to study such cascading failures in networked systems was developed in \cite{Lorenz2009b}, and extended recently to work in more general topologies in~\cite{Burkholz2015}.

In many situations, cascading failures can be influenced by the combination of \em different types \em of interactions between the individual components of the system. This is the case in interbank systems, where banks are exposed to each other via different types of financial obligations (loans, derivative contracts, etc.) (e.g. \cite{KapadiaMay2012} and \cite{MontagnaKok2013}). The bankruptcy of a bank can thus cascade to other banks in non-standard ways. Another example are firms diversifying their activities across different business units, each of which is exposed to cascade risk in its own field of activity. 

An important question we wish to investigate in this article is how diversification across different types of interactions can affect the risk of cascading failures. For that purpose, we study the case of a firm that diversifies its activities across a core-business unit and a subsidiary-business unit. Each business unit is exposed to other firms' business units in the same sector of business activity (either core or subsidiary). This means that a business unit can fail (i.e. go bankrupt) as a result of a cascade of failures (i.e. bankruptcies) in the same sector of business activity.

The question of the structuring of a firm into sub-units has been studied from a different angle in the financial economics literature (e.g. \cite{Lewellen1971} and \cite{MaksimovicPhillips2002}) and often focuses on the efficiency of the allocation of its resources across different industries. Another question that has received some attention is that of whether a firm can diversify the risk of its income streams by operating in different business areas. Namely, \cite{LevySarnat1970}, \cite{SmithSchreiner1969} and \cite{AmihudLev1981} studied how conglomerates can diversify the risk associated with their revenue streams from the perspective of portfolio theory.

\begin{figure}[t]
  \begin{center}
  \includegraphics[scale=0.67,angle=0]{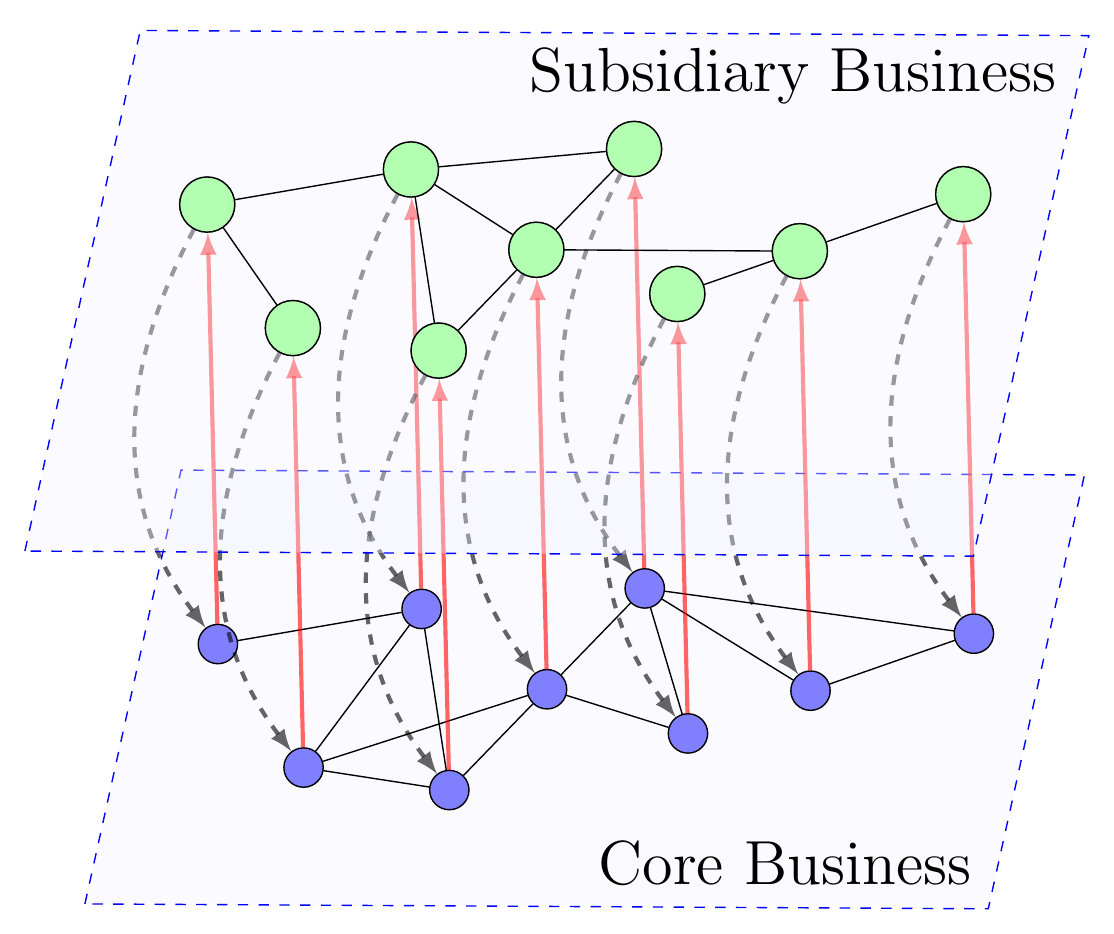}
  \end{center}
  \caption{Illustration of a system with asymmetrically coupled layers. \textit{A failure (or bankruptcy) on the Core Business layer implies a failure on the Subsidiary Business layer. This coupling is illustrated by an inter-layer dependency link (red arrow). On the other hand, a failure on the Subsidiary Business layer only decreases a node's failure threshold on the Core Business layer. This coupling is illustrated using dashed black arrows. The intra-layer links represent business relations or other forms of interactions due to normal business.}}
  \label{fig:illustration}
\end{figure}

Here, we use a complex networks approach and we view the system of firm activities as an interconnected multi-layered network (see~\cite{gao2012, DAgostino2014, kivela2014, Garas2015}). The distinct layers of this network contain individual networks defined by a particular type of interactions according to a given business activity, while the inter-connections between layers allow for cross-layer interactions. 
In this setting we develop a model where failures (i.e. bankruptcies) on two different network layers affect firms asymmetrically: The first layer represents exposures between the firms in the core business while the second layer represents exposures between firms in the subsidiary business. Failure (i.e. bankruptcy) of a firm's core business unit implies failure of its subsidiary business unit, whereas failure of a firm's subsidiary business unit only causes a shock to the firm's resistance threshold in its core business unit (see Fig.~\ref{fig:illustration} for an illustration). We find that when the 
coupling strength from the core to the subsidiary layer is varied only slightly, there is a sharp transition between a safe regime, where there is no cascade of failures, and a catastrophic regime, where there is a full cascade of failures. Moreover, when comparing the two-layer network to the single-layer network formed by aggregating the two layers, we find that cascades can be larger on the two-layer network than on the aggregated one. On the other hand, by varying the strength of the feedback between the two layers, we identify the existence of a regime where the two-layer network is safer than the aggregated one and another regime where the reverse holds. This points to the critical importance of the coupling of the layers when structuring a firm into different business units. Also, dealing with aggregated network data that ignores the fine structure of the coupling between different layers can lead to significant underestimation or overestimation of cascade risk.

The article is structured as follows. In Section \ref{sec:Model}, we describe the two-layer cascade model. In Section \ref{sec:Thermodynamic}, 
we derive a branching process approximation as an approximation for large networks and use it to analyze the aforementioned phenomena. These phenomena are presented in Section \ref{sec:results} where we compare our analytical results with simulations and analyse further the observed phase transitions.
In Section \ref{sec:Conclusion}, we conclude and interpret the consequences of our theoretical investigations for the application to networks of firms that might decide about merging their core and their subsidiary business.

\section{Model}
\label{sec:Model}

We first consider a finite model with $N$ firms. Each firm can be represented by a node $i$ present on each of two different layers: layer $0$ (the \em core-business \em layer) and layer $1$ (the \em subsidiary-business \em layer).
Each layer $l \in \{0,1\}$ 
has a topology represented by an adjacency matrix $G_l$. On each layer $l$, node $i$ can be in one of two states $s^i_l \in \{0, 1\}$, healthy ($s^i_l = 0$) or failed ($s^i_l = 1$). $s^i_0=1$ represents the bankruptcy of firm $i$'s core-business unit, whereas $s^i_1 = 1$ represents the bankruptcy of its subsidiary-business unit. This state is determined by two other variables: 
a node's fragility on a given layer $\phi^i_l$, which accumulates the load a node carries, and its threshold $\theta^i_l$ on that layer, which determines the amount of load it can carry without failing. Whenever the fragility exceeds the threshold $\phi^i_l \geq \theta^i_l$, the node fails on that layer and cannot recover at a later point in time.

On each layer, we assume that a cascade of failures spreads according to the threshold failure mechanism of \cite{Watts2002SimpleModelof}.  Thus a node fails if a sufficient fraction of its neighbors have failed. The fragility of a node $i$ of degree $k_l$ on layer $l$ (i.e. a node with $k_l$ neighbors on layer $l$) can be expressed as 

\begin{equation}
\phi_l^i(k_l) = \frac{1}{k_l} \sum_{j \in \rm{nb_l}(i)}s_l^j = \frac{n_l}{k_l}
\end{equation}
where $\rm{nb_l}(i)$ is the set of nodes in $i$'s neighborhood on layer $l$ and $n_l$ is the number of failed neighbors on layer $l$. This failure mechanism is useful to model a firm diversifying its exposure to failure risk across neighbors: the more neighbors a node has, the less it is exposed to the failure of a single neighbor. A cascade of failures thus starts with an initial fraction of failed nodes. These failures can then spread to their neighbors in discrete time steps. The load $\phi_l^i$ of a node $i$ is thus updated at each time step $t$.
This model has been studied extensively on single-layer networks, in the context of configuration model type random graphs with a given degree distribution \citep{Gleeson2007, Dodds2009, Payne2009, Amini2010, Tessone2012} or trees \citep{Hurd2013}, and has been adapted to financial networks of interbank lending \citep{Gai2010, Battiston2012a, Roukny2013}. In \citet{Burkholz2015} a mesoscopic perspective is added by studying conditional failure probabilities given the degree of node. Generalizations of the model to weighted networks can be found in \citet{Amini2010, Hurd2013} and \citet{Burkholz2015}. 

We extend Watts' model to a multiplex setting with non-symmetric dependence between the layers. The two network layers are related by \em partial dependency links \em (see Fig. \ref{fig:illustration}). 
These are directed links connecting a node on layer $0$ to its alter ego on layer $1$. These links are characterized by weights
$r_{01}, \ r_{10} \in [0,1]$ affecting the size of a shock to a node's threshold on a given layer following a failure on the other layer. Namely, the failure of node $i$ on layer $1$ reduces 
its threshold $\theta^i_0$ on layer $0$ in the following way

\begin{equation}
\theta_0^{i,r} = (1-r_{10}) \theta^i_0,
\end{equation}
while the failure of node $i$ on layer $0$ reduces its threshold $\theta^i_1$ on layer $1$ in the 
following way

\begin{equation}
\theta_1^{i,r} = (1-r_{01}) \theta^i_1.
\end{equation}

In the remainder of this article, we will set $r_{01} = 1$. Thus, the failure of node $i$ on layer $0$ (the core-business layer) automatically leads to its failure on layer $1$ 
(the subsidiary-business layer), since its shocked threshold becomes $\theta_1^{i,r}=0$ and the failure condition $\phi_1^i \geq \theta_1^{i,r}$ is trivially satisfied even in the absence of any failed neighbors on layer $1$. 
On the other hand, we allow $r_{10}$ to take values in $[0,1]$. For $r_{10} = 1$, both layers are fully inter-dependent, meaning that the failure of a node
on one layer implies its failure on the other. In this special case we have normal dependency links between the two layers as introduced in \cite{Buldyrev2010}. 
For $r_{10} = 0$, the failure of a node on layer $1$ does not affect its failure on layer $0$.  For $r_{10} \in (0,1)$, we have an 
asymmetric inter-dependency between layers.                                                                                                       
In the remainder we call $r_{10}$ the \emph{coupling strength} between the two layers.

This models our case of interest, where layer $0$ is interpreted as a firm's activities in a core business, whereas layer $1$ can be interpreted as a firm's activities in a subsidiary
business. The failure of the firm in the subsidiary business does not necessarily imply its failure in the core business, but the loss incurred reduces its ability to withstand the failure of neighbors in the core-business. 
The failure of the firm in the core business however implies its failure in the subsidiary business. It is thus logical to choose the fraction of failed nodes on layer $0$ when the cascade has reached steady state as the appropriate measure of
systemic risk, i.e.

\begin{equation}
\label{eq:X_N^*}
\rho^{*}_N = \lim_{t \rightarrow \infty} \frac{1}{N} \sum^N_{i=1} s_0^i
\end{equation}

\section{Local tree approximation}
\label{sec:Thermodynamic}

In order to make the model analytically tractable, in this section we restrict ourselves to a special class of configuration type multiplex networks. 
For these, we assume that each node is characterized by two degrees, $k_0$, $k_{1}$, and two thresholds,  $\theta_0$, $\theta_{1}$, which can be different on layers $l\in \{0,1\}$. 
These values are drawn independently from the degree distributions $p_l(k_l)$ and the cumulative threshold distributions $F_l(\theta_l)$. 

In the limit of  infinite networks, the clustering coefficient that quantifies the chance that any two neighbors of a given node are also neighbors converges to zero. 
It means that the network is locally tree-like, i.e., it does not contain short cycles.  
This then allows us to develop a branching process approximation for the final fraction of failed nodes $\rho_{l}^{*} := \lim_{N\rightarrow \infty} \rho^{*}_{l,N}$ on each layer $l$ in the limit of infinite network size.  
This approximation is valid for arbitrary degree distributions with finite second moments \citep{Molloy1995,Newman.Strogatz.ea2001Randomgraphswith}.
For our model, the risk measure is $\rho^* = \rho^{*}_0$, i.e., the fraction of failed nodes in layer 0 only. 
However, in order to calculate $\rho^{*}$, we need to calculate both $\rho^{*}_l$, as failures on the two layers are mutually dependent. 

A node fails initially if it has a negative threshold. 
Consequently, $F_0(\theta_{0}=0)$ and $F_1(\theta_{1}=0)$ define, for each layer, the initial fraction of failed nodes. 
These failures can lead to cascades that evolve until the steady states $\rho^{*}_l$ on the two layers are reached.
We can express the $\rho^{*}_l$ as averages with respect to the degree distribution $p_l(k_l)$ which we take as input to our model:
\begin{equation}
  \label{eq:1}
  \rho^{*}_l=\sum_{k_l} p_l(k_l) \mathbb{P}(s_l=1 \vert k_l)
\end{equation}
$\mathbb{P}(s_l=1 \vert k_l)$ is the conditional probability that a node with a given degree $k_l$ on layer $l$ fails in layer $l$. 
In order to calculate this probability, the local tree approximation is essential as it allows us to treat failures of neighbors of a given node as independent events that happen with a failure probability $\pi_{l}^{*}$. 
Consequently, the number $n_{l}$ of failed neighbors of a node of degree $k_l$ in layer $l$ follows a binomial distribution so that
$n_l$ neighbors are failed with probability 
\begin{equation}
 B(n_l, k_l, \pi^{*}_l) := \binom{k_l}{n_l} \left(1-\pi^{*}_l\right)^{k_l-n_l} \left( \pi^{*}_l \right)^{n_l}.
\label{eq:2}
\end{equation}
This allows us to express $\mathbb{P}(s_l=1 \vert k_l)$ as: 
\begin{equation}
  \label{eq:pfail}
   \mathbb{P}(s_l=1 \vert k_l) = \sum^{k_l}_{n_l = 0} B(n_l, k_l, \pi^{*}_l) \mathbb{P}\left(s_l=1\vert k_l, n_l, \rho^*_{s,1-l}\right),
\end{equation}
where $\mathbb{P}(s_l=1\vert k_l, n_l, \rho^*_{s,1-l})$ is the probability that a node fails in layer $l$ given that exactly $n_{l}$ of its $k_{l}$ neighbors failed. 
This failure results from the fact that the fraction of failed nodes $n_{l}/k_{l}$ exceeded the threshold.
This was the \emph{original} threshold $\theta_{l}$ with a probability $(1-\rho^*_{s,1-l})$ because the node did not fail in layer $(1-l)$. 
Or it was the \emph{reduced} threshold $\theta_{l}(1-r_{(1-l)(l)})$ with a probability $\rho^*_{s,1-l}$ because the node failed in layer $(1-l)$ before.
Consequently, we have:
\begin{align}\label{eq:pfail2}
\mathbb{P}\left(s_l=1\vert k_l, n_l, \rho^*_{s,1-l}\right) =  \left(1-\rho^*_{s,1-l}\right) F_l\left(\frac{n_l}{k_l} \right) + \rho^*_{s,1-l} F_l\left(c_l \frac{n_l}{k_l} \right).
\end{align}
$F_l\left({n_l}/{k_l}\right)$ is the  probability that the original threshold is exceeded, whereas 
$F_l\left(c_l {n_l}/{k_l} \right)$ is the probability that the reduced threshold is exceeded,  where $c_0 := 1/(1-r_{10})$ and $c_1 := \infty$.

Note that $\rho^*_{s,1-l}$ differs from $\rho_{1-l}^{*}$ used in Eqn. \eqref{eq:1}. It gives us only the conditional probability that the node has failed in layer $(1-l)$, given that it has not failed in layer $l$. 
It depends on a neighbor's failure probability on the other layer, $\pi^{*}_{1-l}$, via
\begin{align}\label{eq:singlefail}
 \rho^*_{s,1-l} = \sum_{k_{1-l}} p_{1-l}(k_{1-l}) \sum^{k_{1-l}}_{n_l = 0} B(n_l, k_{1-l}, \pi^{*}_{1-l})\  F_{1-l}\left(\frac{n_{1-l}}{k_{1-l}}\right).
\end{align}
Eqn. \eqref{eq:singlefail} has the same  structure as the branching process approximation for the fraction of failed nodes on \emph{single} layers \citep{Gleeson2007}.
It also has a structure similar to Eqs. \eqref{eq:1}-\eqref{eq:pfail}, with the only difference of a simpler response function, i.e.,  $\mathbb{P}(s_l=1\vert k_l, n_l, \rho^*_{s,1-l})=F_{1-l}(n_{1-l}/k_{1-l})$, which is the probability that the fraction of failed neighbors exceeds the original threshold in layer $(1-l)$.

In order to compute Eqn. \eqref{eq:pfail}, we need to  know the failure probability  $\pi_l^*$ of a neighbor in layer $l$. 
To achieve this, we iteratively solve a system of coupled fixed-point equations for the probabilities $\pi_l^*$, defined as
\begin{align}\label{eq:fixp}
 \pi^{*}_l = L_l\left( \pi^{*}_l\right) := \frac{1}{z_{l}} \sum_{k_l} {p_l(k_l)k_l} \sum^{k_l-1}_{n_l = 0} B(n_l, k_l-1, \pi^{*}_l) \mathbb{P}\left(s_l = 1 \vert k_l, n_l, \rho^*_{s,1-l}\right)
\end{align}
where $z_l = \sum_{k_l} p_l(k_l) k_l$ and $\mathbb{P}(s_l = 1 \vert k_l, n_l, \rho^*_{s,1-l})$ is defined by Eqn. \eqref{eq:pfail2}

We note again similarities between Eqn.~\eqref{eq:fixp} that describes the failure probability of a \emph{neighbor} and Eqs.~\eqref{eq:1}, \eqref{eq:pfail2} that describe the failure probability of a \emph{node}, but also two differences: 

a) In Eqn. \eqref{eq:fixp} the degree distribution of a \emph{neighbor} follows ${p_l(k_l)k_l}/{z_l}$ instead of $p_l(k_l)$ for a \emph{node}  \citep{Newman2010}. It is proportional to the degree $k_l$, since every link of a neighbor increases the probability of the neighbor to be connected to the node under consideration. 

b) The binomial distribution in Eqn. \eqref{eq:fixp} depends on $k_{l}-1$ instead of $k_{l}$ because we have to take into account the second-order neighborhood of the node under consideration. 
We take the failure probability of a neighbor as input to calculate the failure probability of the node under consideration as in Eqs. \eqref{eq:1}, \eqref{eq:pfail2}. 
Therefore, $\pi^{*}_{l}$ is conditioned on the event that the node under consideration has not failed before the neighbor.
Only $k_l -1$ neighbors of the neighbor with degree $k_l$ can have possibly failed before, because the node under consideration is a neighbor of this neighbor.

\section{Results}
\label{sec:results}

\subsection{Comparison with computer simulations}
\label{sec:comp-with-comp}

We now compare our numerical solution of the fixed point equations \eqref{eq:fixp} with computer simulations, using the illustrative case of an uncorrelated, two-layer Erd\H{o}s-R\'enyi network \citep{ErdosRenyi}. For the computer simulations, we implement the time-dependent model as described in Sect. \ref{sec:Model}, i.e. we simulate the evolution of cascades until they reach the steady state. 
The network size is $N=10,000$ for each layer. Further, we sample over $100$ network realizations for every initial condition. 
The degree distributions $p_{l}(k_{l})$ on each layer are approximately Poisson distributions
\begin{equation}
  \label{eq:3}
  p_{l}(k_{l})= \frac{z_{l}^{k_{l}}}{k_{l}!}\ e^{-z_{l}}
\end{equation}
with identical mean degrees $z_0 = z_1$. The thresholds are normally distributed, i.e. $\theta_0 \sim \mathcal{N}(\mu_0,\sigma^2_0)$ and $\theta_1 \sim \mathcal{N}(\mu_1,\sigma^2_1)$ with \emph{different} parameters $\mu_{l}$, $\sigma_{l}$. In our case, we fix the threshold distribution on layer 0 
to parameters $\mu_0=0.3$ and $\sigma_0=0.1$, which ensures that the failure probability for the nodes in an uncoupled network would be very small, i.e., $0.0045$.  
We vary the parameters $\mu_1$ and $\sigma_1$ of layer 1, to study the emergence of large cascades on layer 0 because of the coupling with strength $r_{10}$.

\begin{figure}[t]
  \centerline{
\includegraphics[scale=0.3]{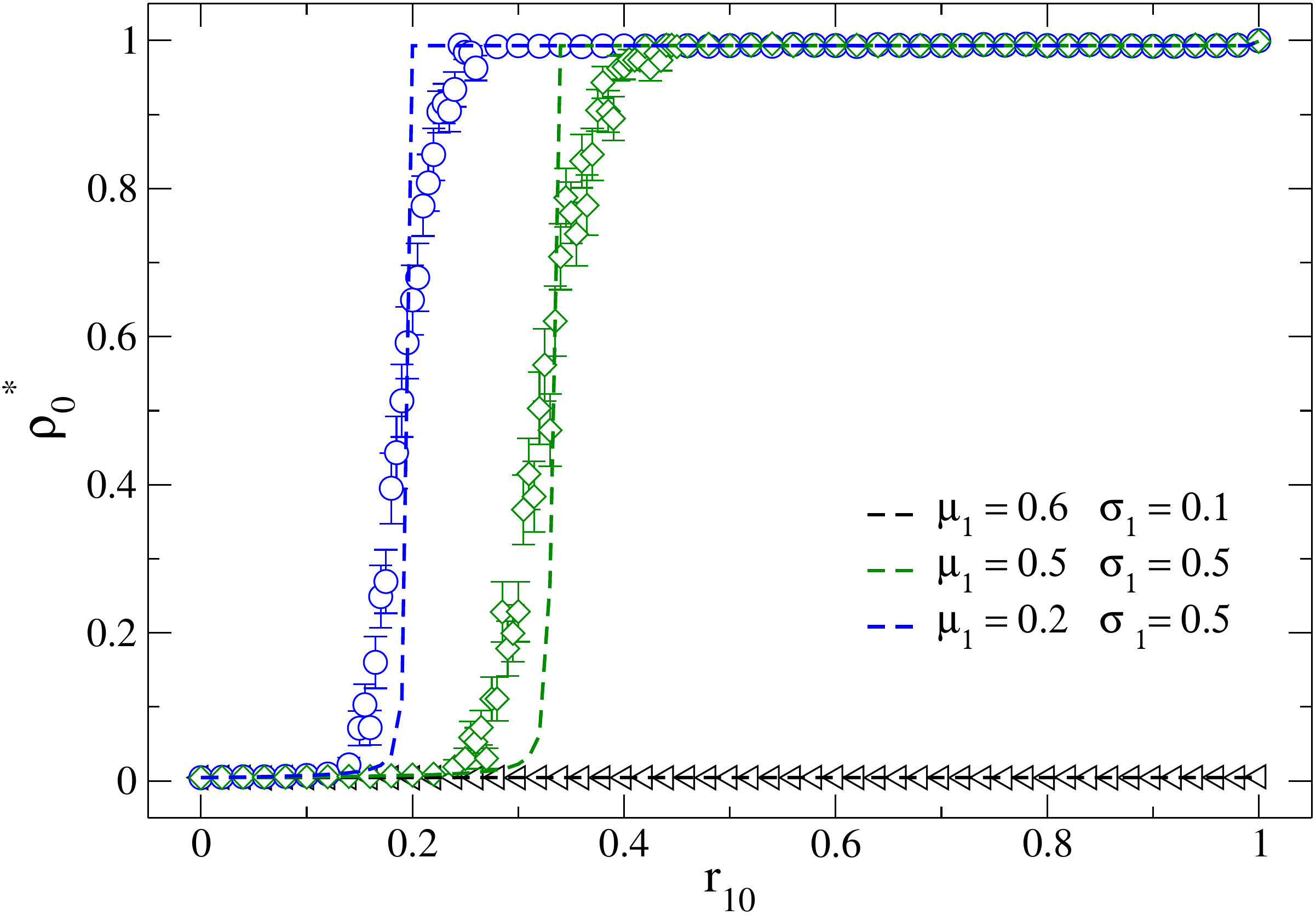}
  }
  \caption{Sharp Regime Change in Threshold-Feedback Model: \textit{$\rho^*_0$ for different combinations of threshold distributions, as the coupling strength $r_{10}$ is varied.  The threshold distribution on the core-business layer is set to $\mathcal{N}(0.3,0.1^2)$ while the threshold distribution on the subsidiary-business layer is $\mathcal{N}(0.6,0.1^2)$ (black curve), $\mathcal{N}(0.5,0.5^2)$ (green curve) and $\mathcal{N}(0.2,0.5^2)$ (blue curve). The two layers are independent Erd\H{o}s-R\'enyi networks with mean degrees $z_0=z_1=5$. The dotted lines are the curves predicted by our analytics, while with the open symbols we show simulation results on Erd\H{o}s-R\'enyi networks with 10000 nodes where each point is averaged over 100 realizations. The error bars indicate the size of the standard error.}}
  \label{fig:tipping_versus_r21}
\end{figure}

Figure \ref{fig:tipping_versus_r21} demonstrates that a small variation in the coupling strength $r_{10}$ may result in a rapid shift from a regime with almost no failures ($\rho^*_0 \approx 0$) to a regime of complete system failure ($\rho^*_0 \approx 1$). 
The coupling strength at the onset of this regime shift, $r_{10}^{c}$, depends on the parameters of the threshold distribution on layer 1 (i.e. $\mu_{1}$ and $\sigma_{1}$). 
Namely, $r_{10}^{c}$ is increasing in $\mu_{1}$.
For certain parameter constellations (e.g. $\mu_{1}=0.6$ and $\sigma_{2}=0.1$), we also find \emph{no} failure cascades at all, which will be further discussed in Fig. \ref{fig:mu2_r21_plot_mu1_03_sigma1_01_sigma2_03}. 
Finally, we note the excellent match between the numerical and the simulation results. 
The differences in the slope result from the fact that we have simulated finite networks, whereas the fixed point equations holds for infinite networks. 

\subsection{Impact of the coupling strength}
\label{sec:impact-coupl-strengt}

\begin{figure}[t]
\centering
  \begin{tabular}{@{}cccc@{}}
 & \bf $\rho^*_0$ & \bf $\rho^*_1$ & \bf $\rho^*_{\text{agg}}$\\
 \bf $r_{10} = 0$  &
\includegraphics[valign=m,width=.23\textwidth]{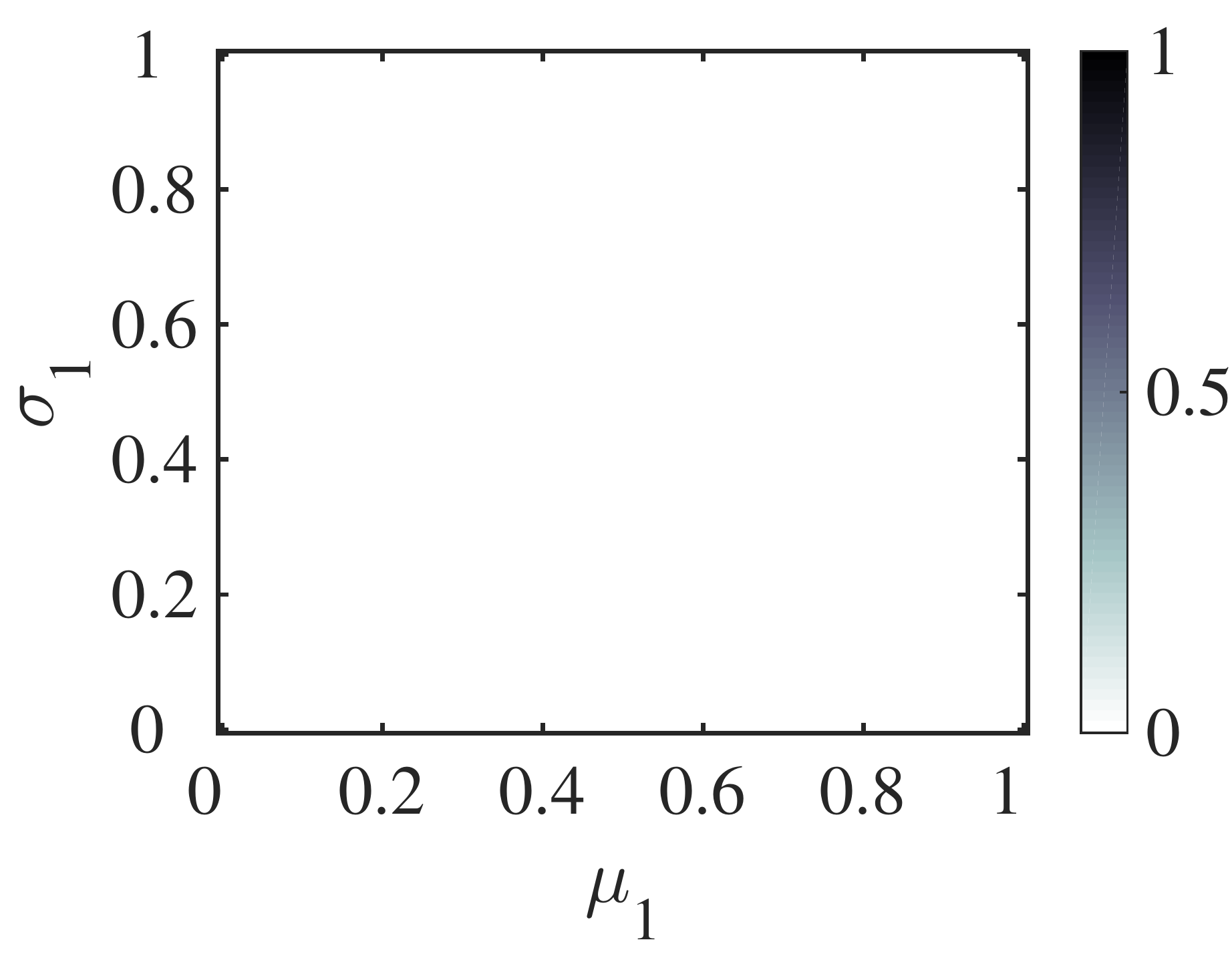} &
    \includegraphics[valign=m,width=.23\textwidth]{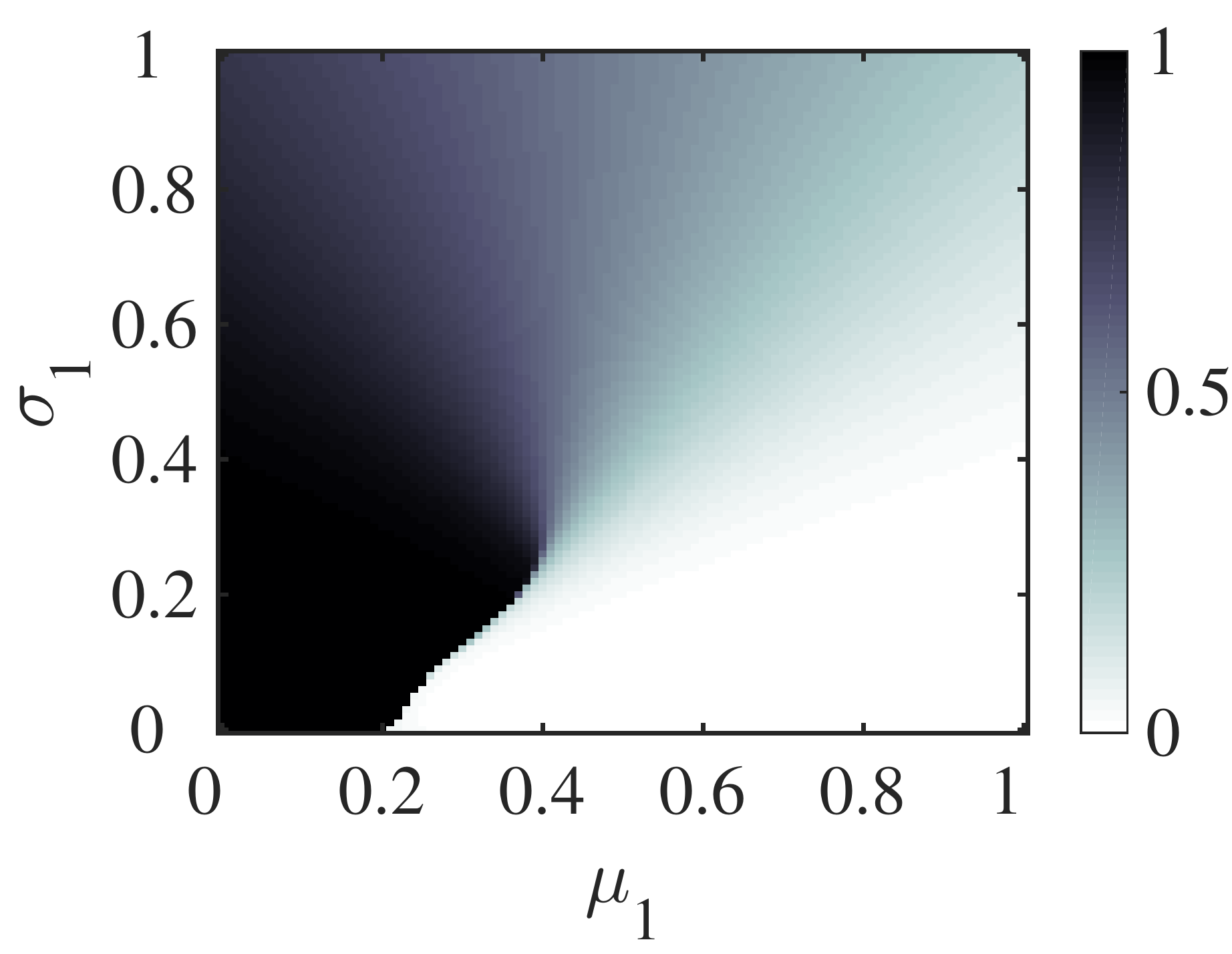} &
    \includegraphics[valign=m,width=.23\textwidth]{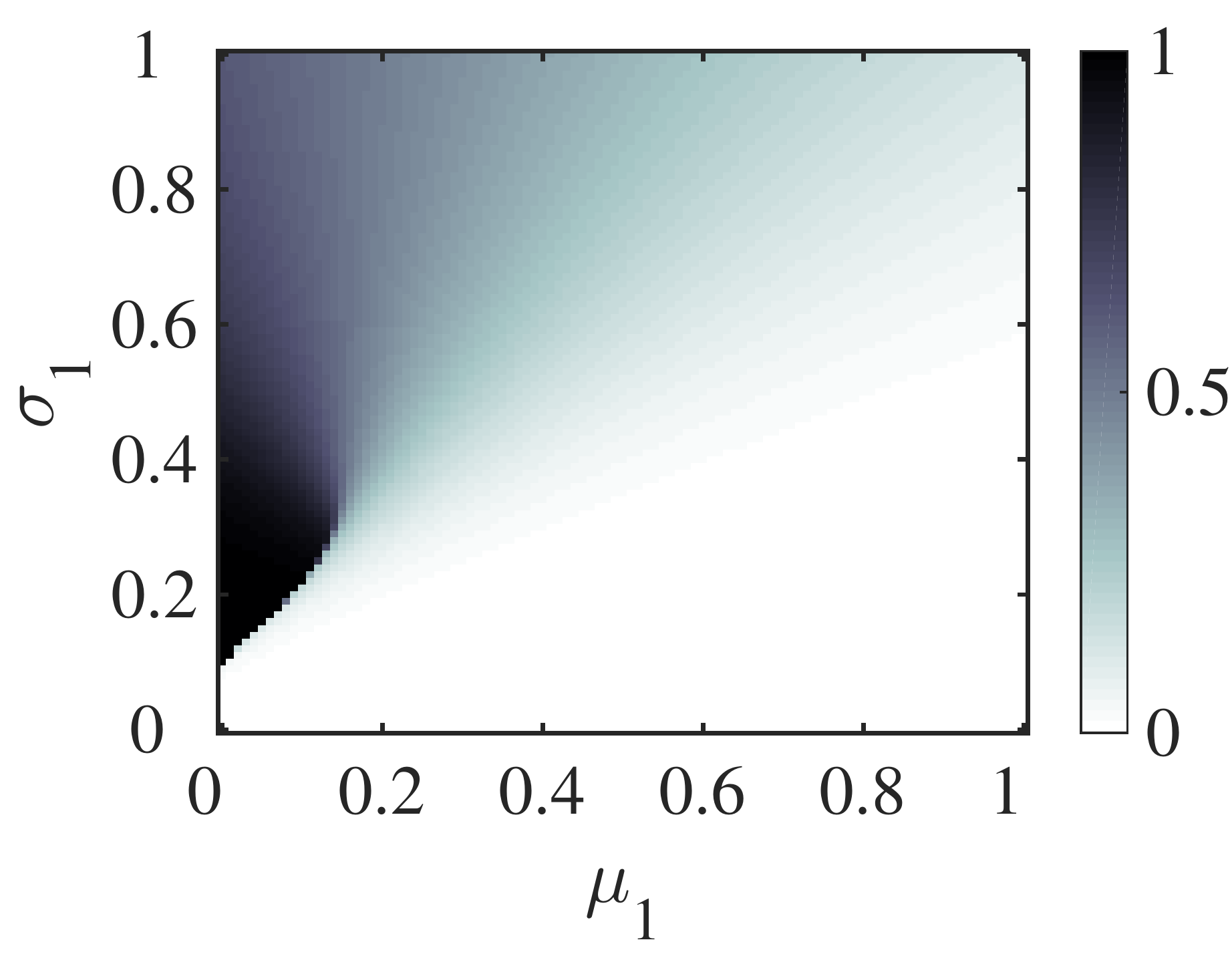}   \\

   \bf $r_{10} = 0.2$ &
    \includegraphics[valign=m,width=.23\textwidth]{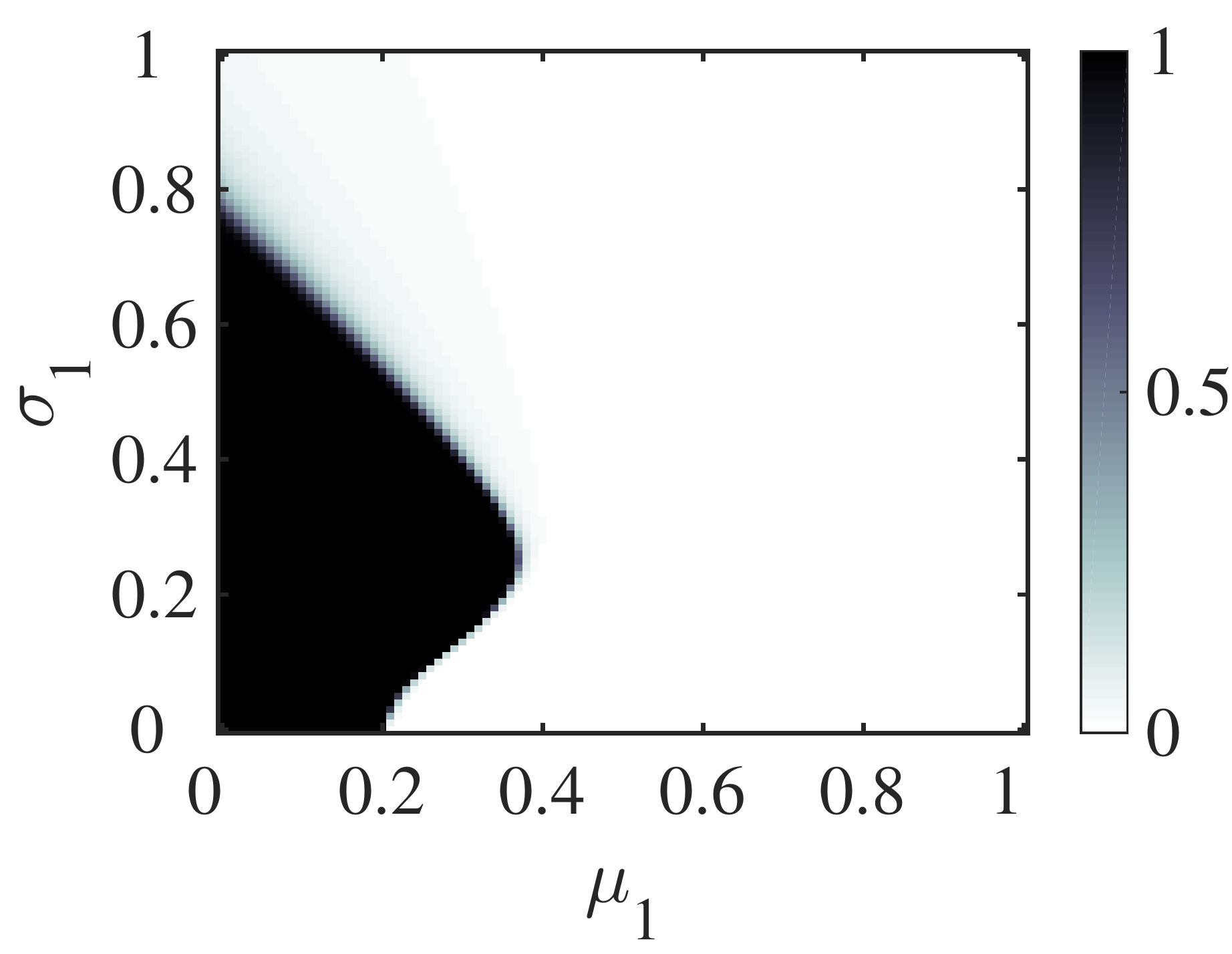} &
    \includegraphics[valign=m,width=.23\textwidth]{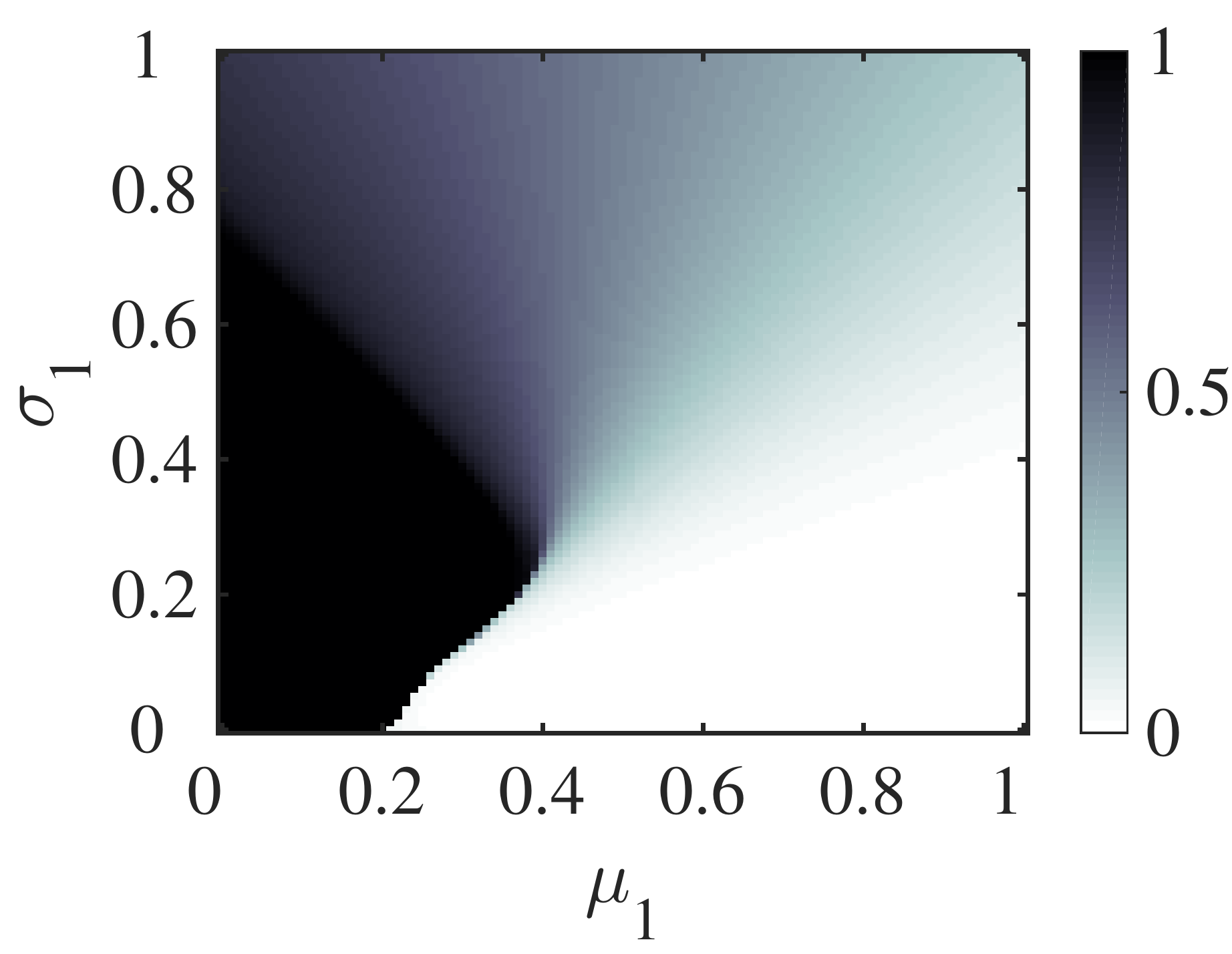} &
    \includegraphics[valign=m,width=.23\textwidth]{Fig3agg}   \\
    \bf $r_{10} = 0.4$ &
    \includegraphics[valign=m,width=.23\textwidth]{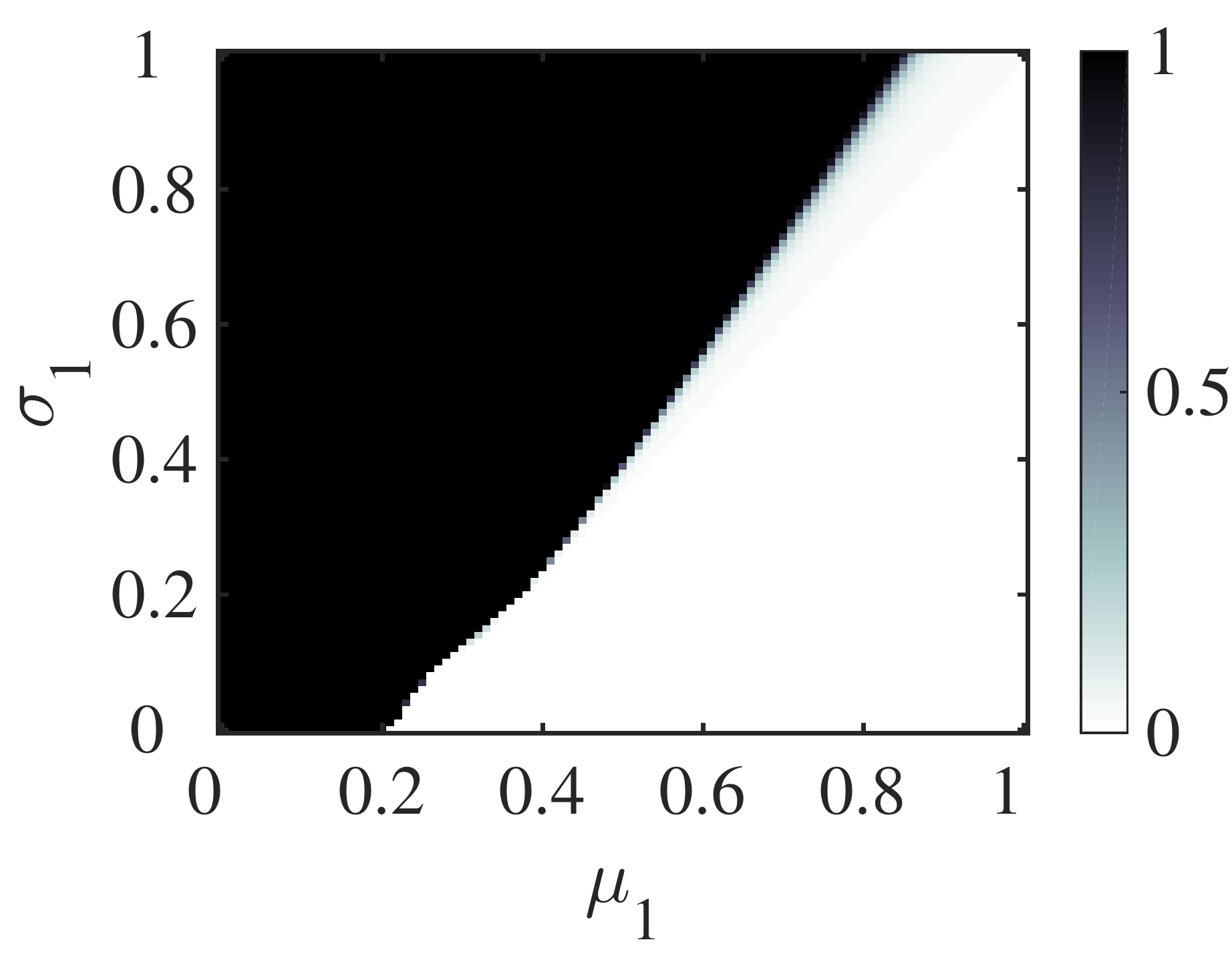} &
    \includegraphics[valign=m,width=.23\textwidth]{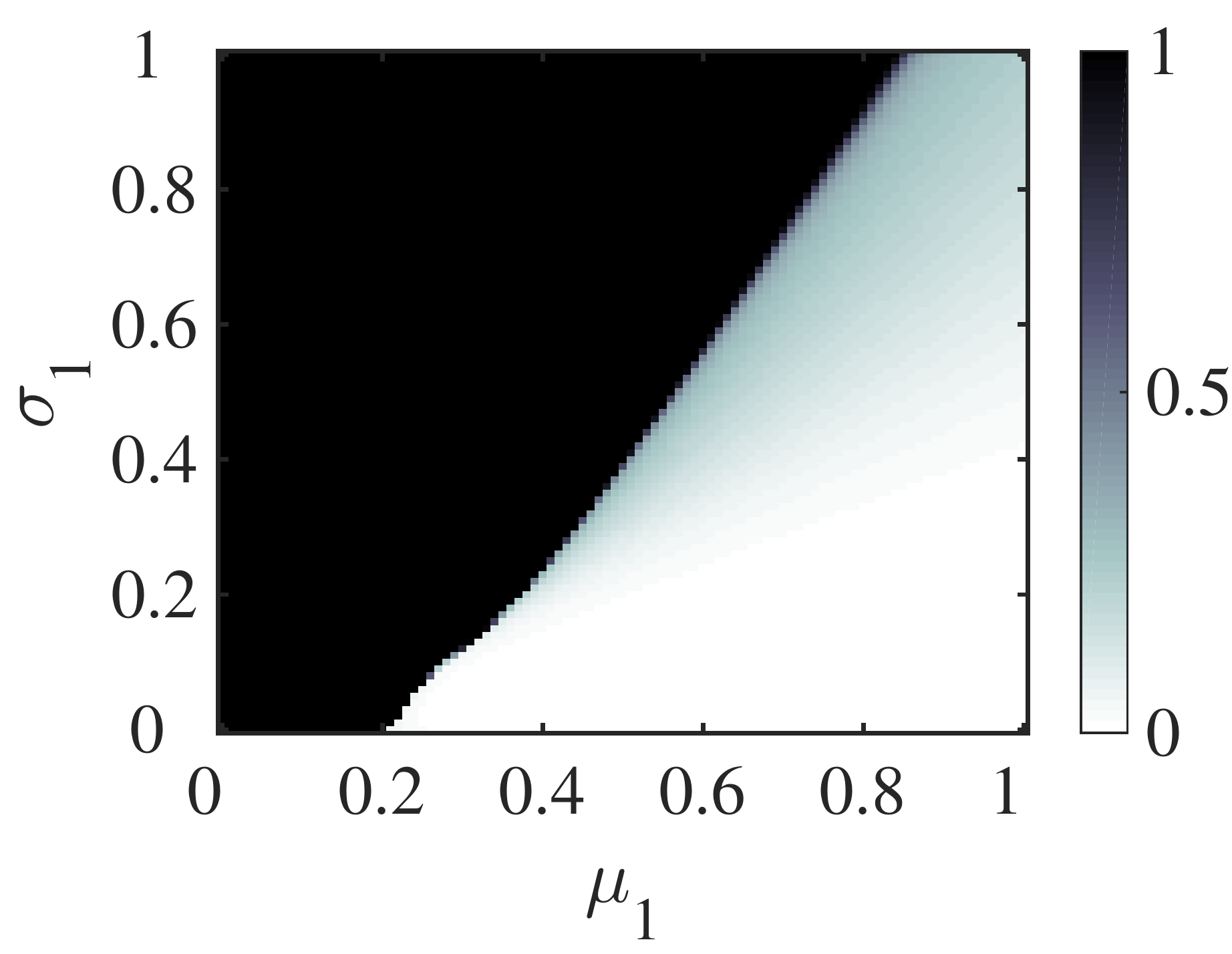} &
    \includegraphics[valign=m,width=.23\textwidth]{Fig3agg}   \\
    \bf $r_{10} = 0.8$  &
    \includegraphics[valign=m,width=.23\textwidth]{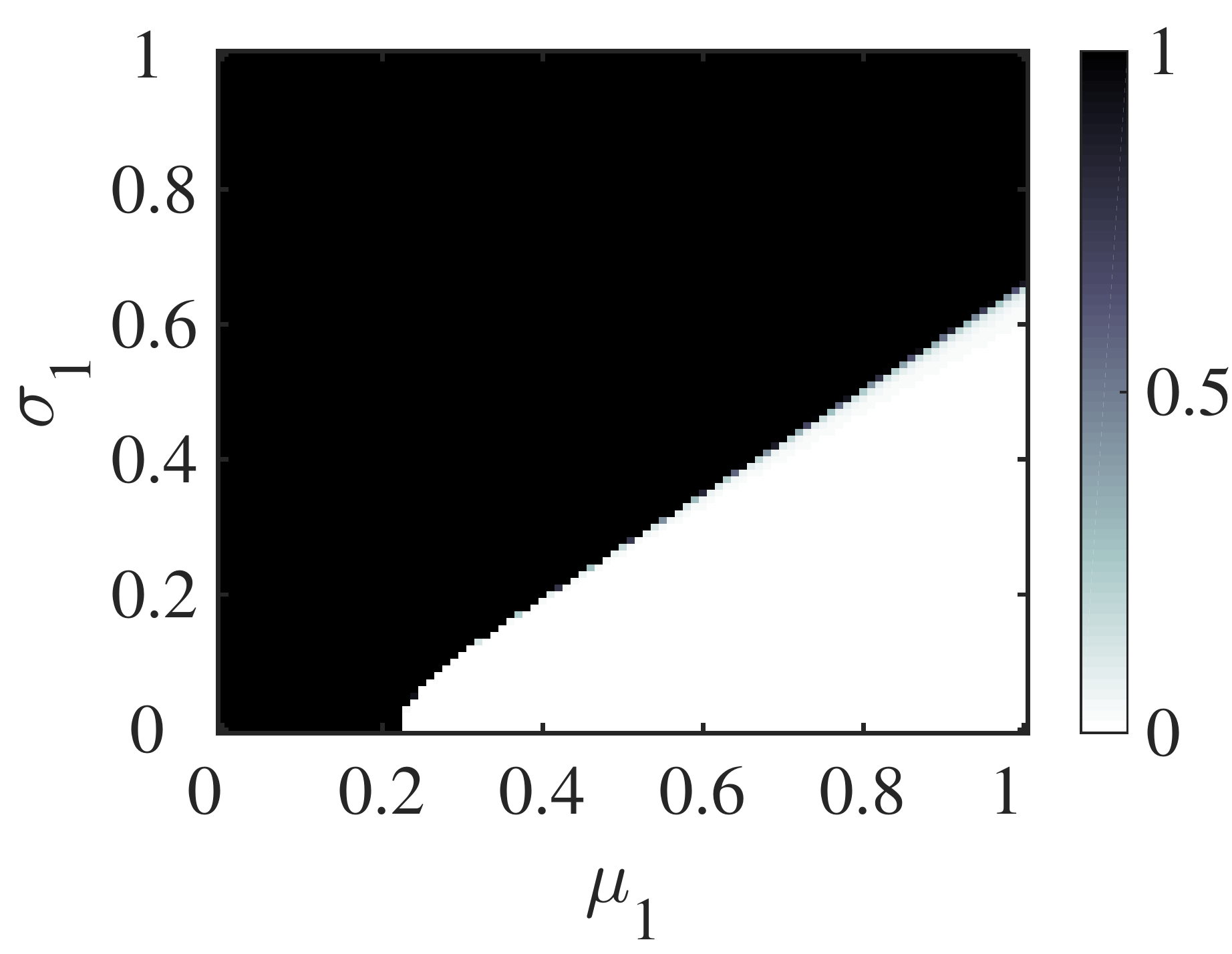} &
    \includegraphics[valign=m,width=.23\textwidth]{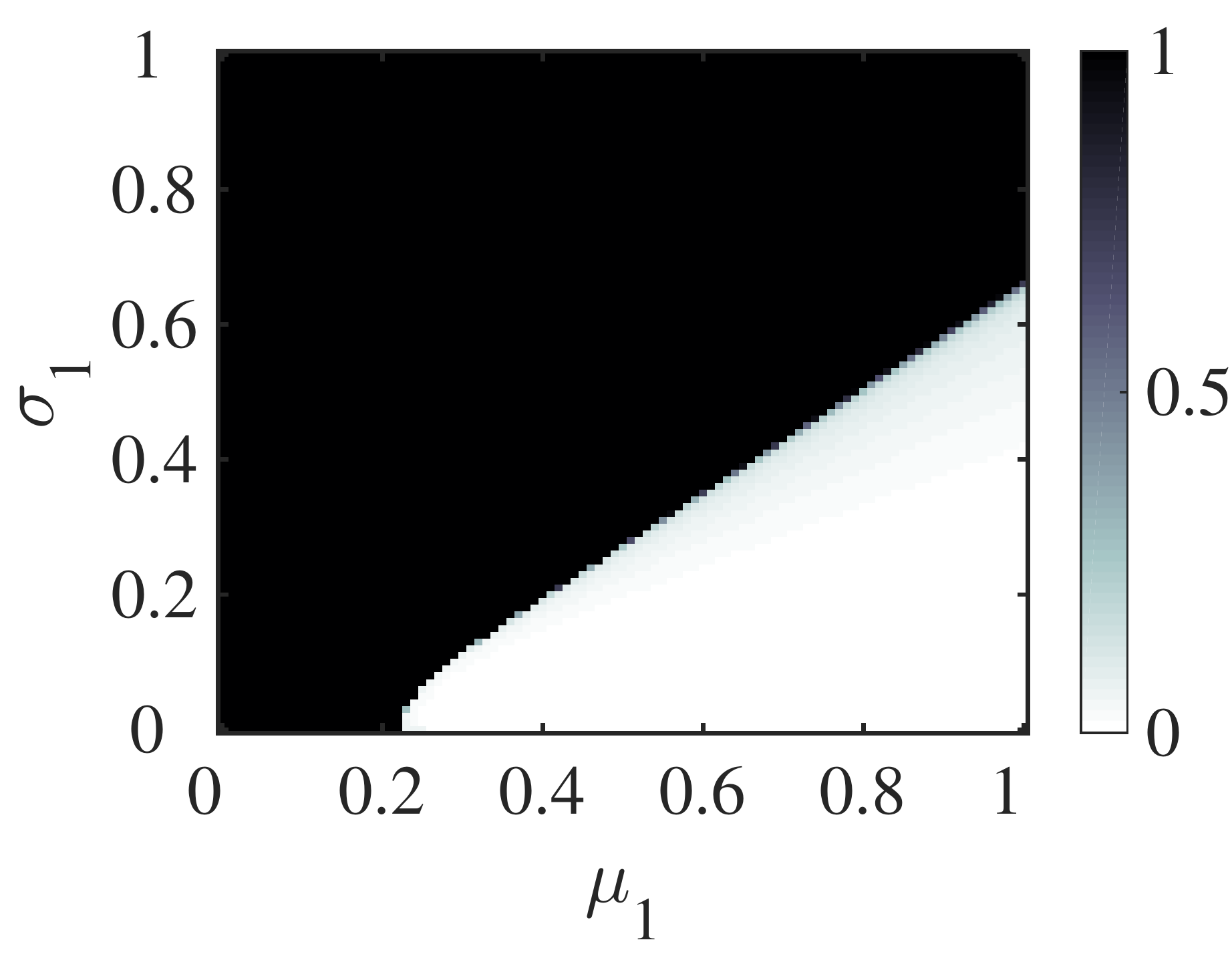} &
    \includegraphics[valign=m,width=.23\textwidth]{Fig3agg}   \\
    \bf $r_{10} = 1.0$  &
    \includegraphics[valign=m,width=.23\textwidth]{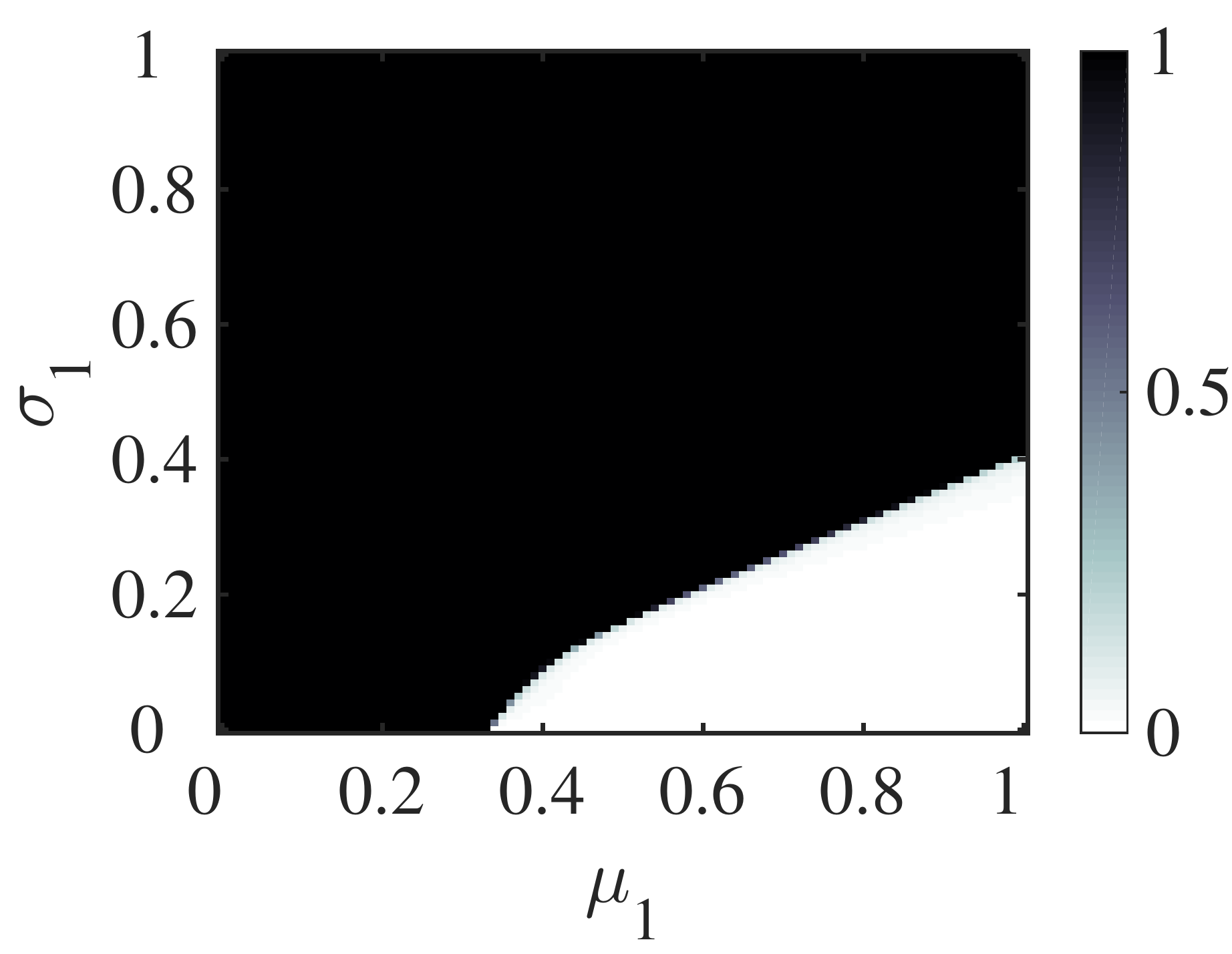} &
    \includegraphics[valign=m,width=.23\textwidth]{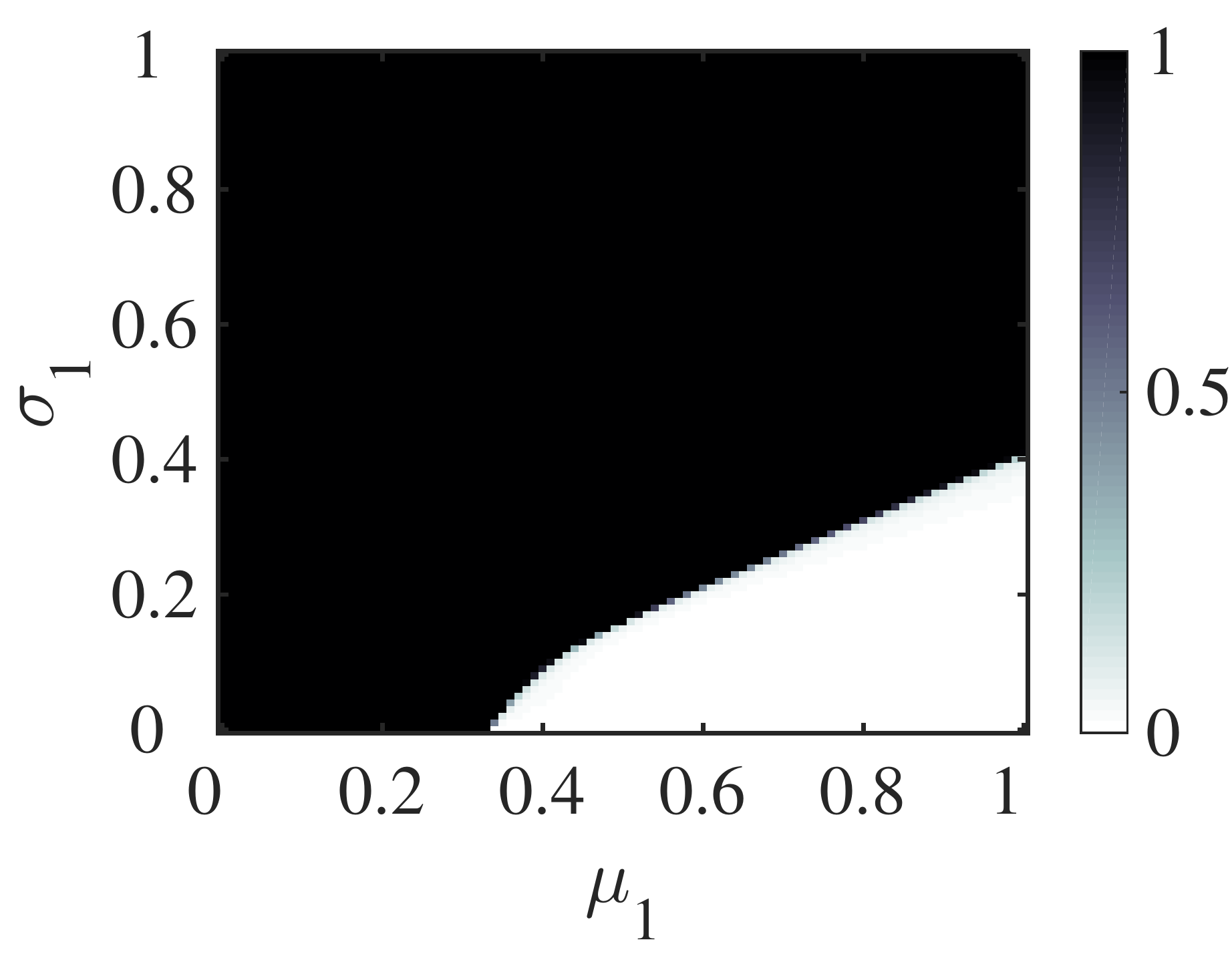} &
    \includegraphics[valign=m,width=.23\textwidth]{Fig3agg}   
    \end{tabular}
  \caption{Sharp Regime Change in Threshold-Feedback Model: \textit{$\rho^*_0$ and $\rho^*_1$ for $\mu_0=0.3$, $\sigma_0=0.1$ and $(\mu_1,\sigma_1) \in [0,1]^2$ as $r_{10}$ is varied. 
  The two layers are independent Erd\H{o}s-R\'enyi networks with mean degrees $z_0=z_1=5$ and thresholds on each layer are independently distributed, i.e. $\theta_0 \sim \mathcal{N}(0.3,0.1^2)$ and $\theta_1 \sim \mathcal{N}(\mu_1,\sigma^2_1)$,
   where $(\mu_1,\sigma_1) \in [0,1]^2$. $\rho^*_{agg}$ is the fraction of failures on the aggregated network, i.e. a network where $k_{agg} = k_0 + k_1$ and $\theta_{agg} = \theta_0 + \theta_1$.}}
   \label{fig:z1_5_z2_5_muc_03_sigc_01_agg}
\end{figure}

To gain a broader understanding of how the coupling between the two layers can cause such a rapid transition from a low-risk regime (i.e. $\rho^*_0 \approx 0$) to a catastrophic one (i.e. $\rho^*_0 \approx 1$), we calculate the $(\mu_{1},\sigma_{1})$ phase diagram for various coupling strengths.
The results are shown in Fig. \ref{fig:z1_5_z2_5_muc_03_sigc_01_agg}, where dark areas indicate parameter constellations with a very high fraction of failed nodes. 
The left column shows the measure of systemic risk $\rho_{0}^{*}$, which is equivalent to the fraction of failed nodes in layer 0, whereas the middle column shows the corresponding fraction of failed nodes in layer 1.

As a reference case, the right column shows a single-layer network, which we constructed in order to put the results for the two-layer network into perspective.
Nodes in this combined single-layer network have a degree $k_{agg} = k_0 + k_1$ and a threshold $\theta_{agg} = \theta_0 + \theta_1$, i.e. we simply sum up the two values per node that were distributed on the two layers before. 
The degree distribution can be found using the convolution of $p_0(k_0)$ and $p_1(k_1)$, i.e. $p_{agg} (k_{agg})= (p_0 \ast p_1) (k_{agg})$. 
The latter can be shown to be a Poisson distribution with mean degree $z_{agg} = z_0 + z_1$, which is the degree distribution of an infinite Erd\H{o}s-R\'enyi network.
The threshold distribution for the combined layers can be found by taking the convolution of the probability density functions of $\theta_0$ and $\theta_1$.
This yields $\theta_{agg} \sim \mathcal{N}(\mu_{agg},\sigma^2_{agg})$, where $\mu_{agg}=\mu_0+\mu_1$ and $\sigma_{agg} = \sqrt{\sigma_0^2 + \sigma_1^2}$. 

Our reference case is motivated by two considerations: (a) We want to estimate the error made if a multiplex network is approximated by a single layer network, i.e. the properties of the different layers are simply aggregated in one layer. (b) For the application scenario at hand, namely the management decision of firms to merge their core and subsidiary business units into one business, we want to understand the impact on the resulting risk exposure. 
To calculate the phase diagram, we assume that all firms make the same decision, which for example could be motivated by herding behavior.

Since we vary only $(\mu_1,\sigma_1)$, we plot all phase diagrams with respect to these two parameters. 
Because the combined layer network no longer contains the coupling strength $r_{10}$, the respective phase plots do not change by varying $r_{10}$.
They are merely repeated for the purpose of comparison with the other columns.
We hypothesize that nodes in the combined layer network have a smaller failure probability compared to the two-layer network because their threshold is much higher. 
Their degree is also larger, which implies that the risk is better diversified among the neighbors. 
On the other hand, because of the larger degrees, there is a higher connectivity in the combined layer network. 
This has the potential to amplify small cascades more than in the less connected separate layers. 
We will investigate, by means of numerical solutions of the fixed point equations, which of these antagonistic effects may dominate in a given parameter region. 

By comparing the first and third columns in Fig. \ref{fig:z1_5_z2_5_muc_03_sigc_01_agg}, we see a different risk profile for small and for large coupling strengths. 
For small values of $r_{10}$ the cascades on layer 1 cannot propagate to layer 0, therefore we do not observe any systemic risk. 
This is different for the combined layer network, where large cascades can occur for a small range of parameters.  

The picture is inverted for larger values of $r_{10}$. 
Here we find, by increasing $r_{10}$, an increasing region of high systemic risk that is driven by the mutual amplification of cascades between the two layers. 
This leads to a very sharp phase transition, i.e. a clear separation of regions with complete breakdown and regions with no breakdown.
We note that this differs from the observation for the combined layer network, where the phase transition can also be observed.
However, there are extended regions where the systemic risk is at intermediate levels, as indicated by the gray areas. 

\begin{figure}[tb]
\centering
  \begin{tabular}{@{}ccc@{}}
  \bf $\rho^*_0$ & \bf $\rho^*_1$ & \bf $\rho^*_{\text{agg}}$\\
 
\includegraphics[valign=m,width=.3\textwidth]{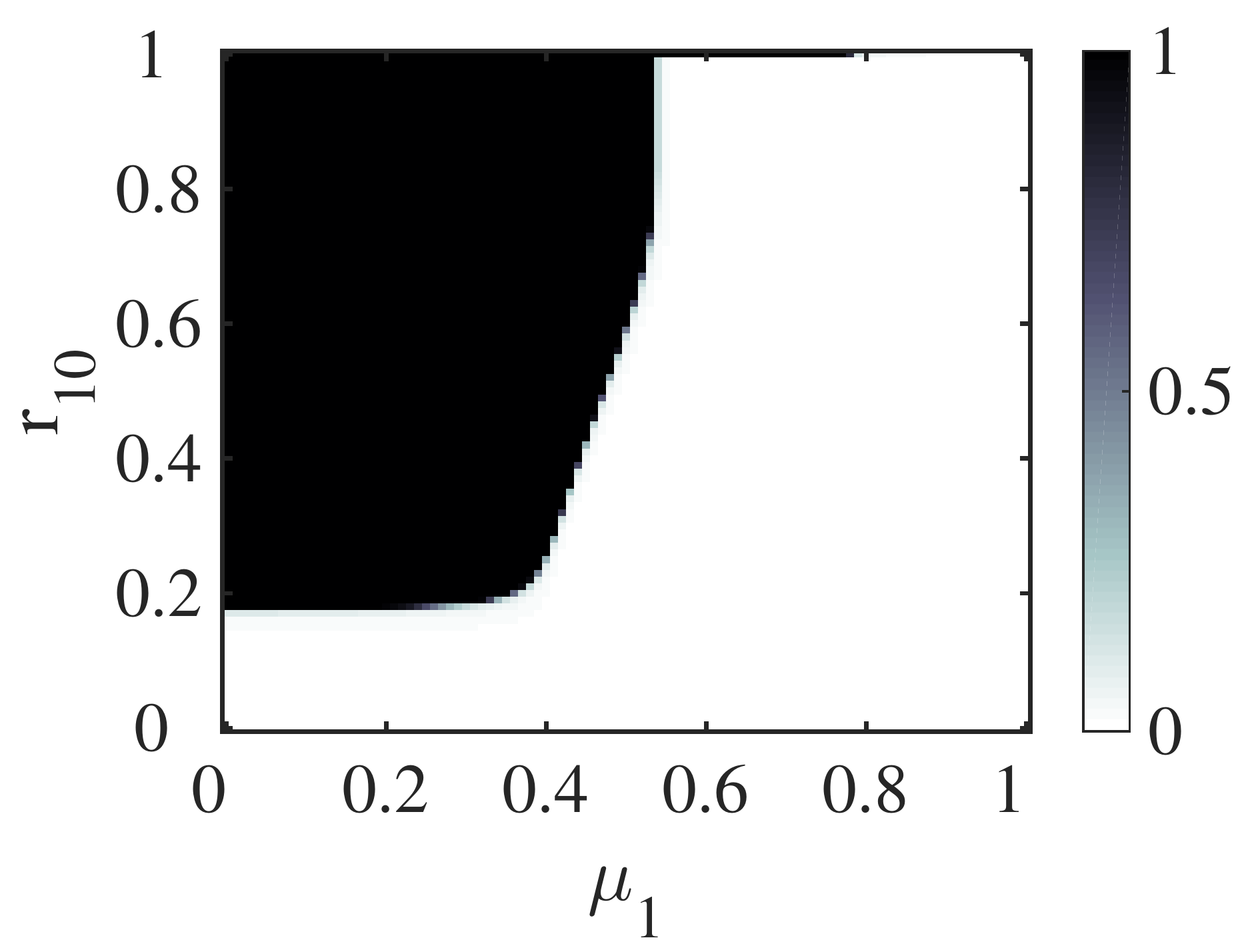} &
    \includegraphics[valign=m,width=.3\textwidth]{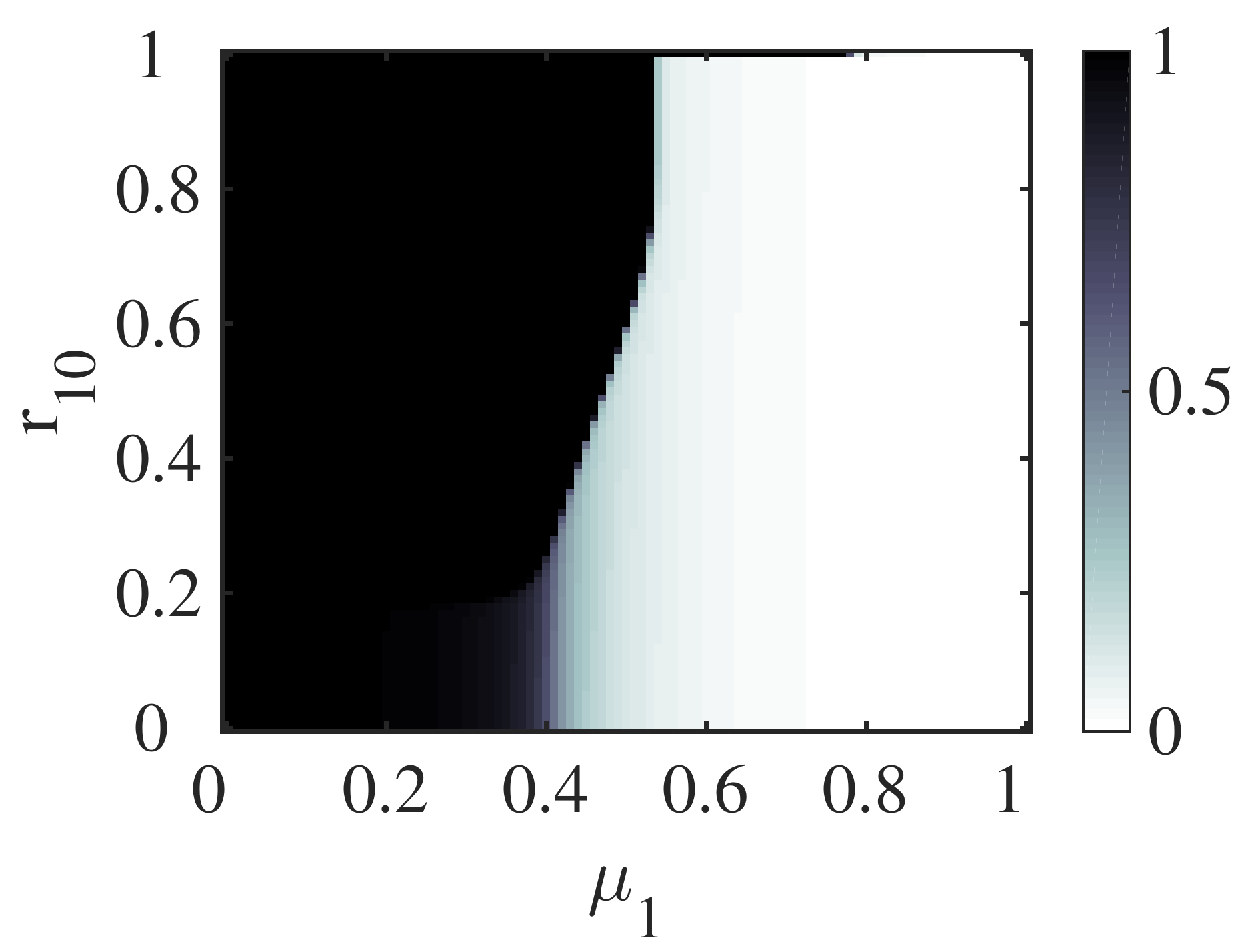} &
    \includegraphics[valign=m,width=.3\textwidth]{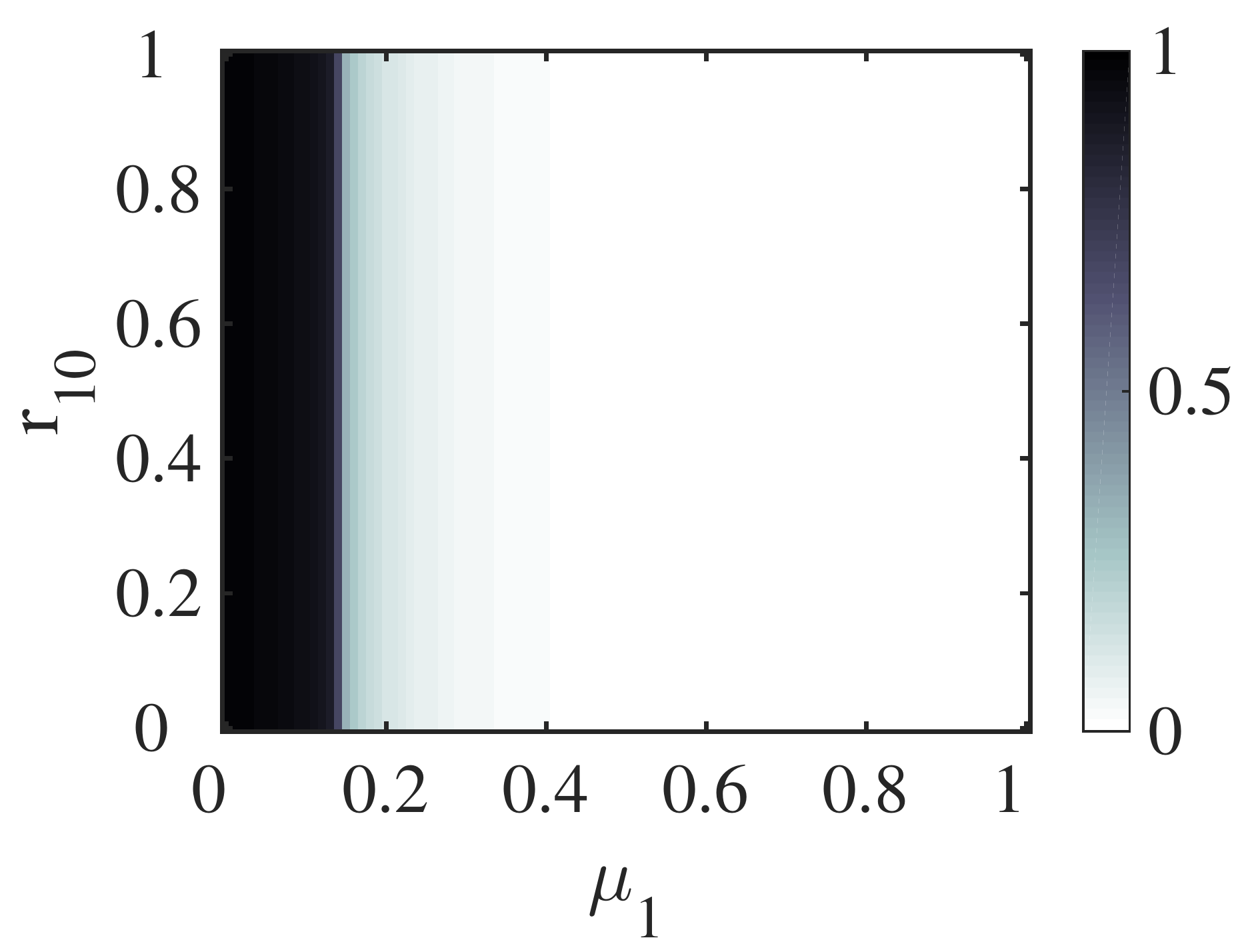}   \\
    \end{tabular}
  \caption{Sharp Regime Change in Threshold-Feedback Model: \textit{$\rho^*_0$ and $\rho^*_1$ for $\mu_0=0.3$, $\sigma_0=0.1$ and $\sigma_1=0.3$ as $\mu_1$ and $r_{10}$ are varied. 
  The two layers are independent Erd\H{o}s-R\'enyi networks
  with mean degrees $z_0=z_1=5$ and the thresholds on each layer are independently distributed, i.e. $\theta_0 \sim \mathcal{N}(0.3,0.1^2)$ and $\theta_1 \sim \mathcal{N}(\mu_1,0.3^2)$,
   where $\mu_1 \in [0,1]$. $\rho^*_{agg}$ is the fraction of failures on the aggregated network, i.e. a network where $k_{agg} = k_0 + k_1$ and $\theta_{agg} = \theta_0 + \theta_1$.}}
   \label{fig:mu2_r21_plot_mu1_03_sigma1_01_sigma2_03}
\end{figure}

We also wish to see how the \emph{onset} of systemic risk on layer 0 depends on the coupling strength $r_{10}$.  
Thus, in addition to fixing $\mu_{0}$ and $\sigma_{0}$, we set $\sigma_{1}$ to a small value and plot the phase diagram with respect to $(\mu_{1},r_{10})$.  
Fig. \ref{fig:mu2_r21_plot_mu1_03_sigma1_01_sigma2_03}, for the same columns as in Fig. \ref{fig:z1_5_z2_5_muc_03_sigc_01_agg}, shows that there is indeed a critical value $r_{10}^{c}$ which is independent of $\mu_{1}$. 
Below $r_{10}^{c}$, we do not observe any systemic risk in layer 0, whereas in the combined layer network, we find a considerable systemic risk for small values of $\mu_{1}$. 
Above $r_{10}^{c}$, we see in the two-layer network a sharp phase transition between full and no systemic risk that depends on a critical value $\mu_{1}^{c}$. 
Consequently, we study the transition line $r_{10}(\mu_{1})$ in the following subsection. 

\subsection{Scaling behavior}
\label{sec:scaling}

\begin{figure}[t]
\centering
  \begin{tabular}{@{}ccc@{}}
  \bf Single layer $\rho^*_{\text{single}}$ & \bf Slope $s_1$ & \bf $\rho^*_{0}$ for $r_{10} = 0.3$\\
 
    \includegraphics[valign=m,width=.32\textwidth]{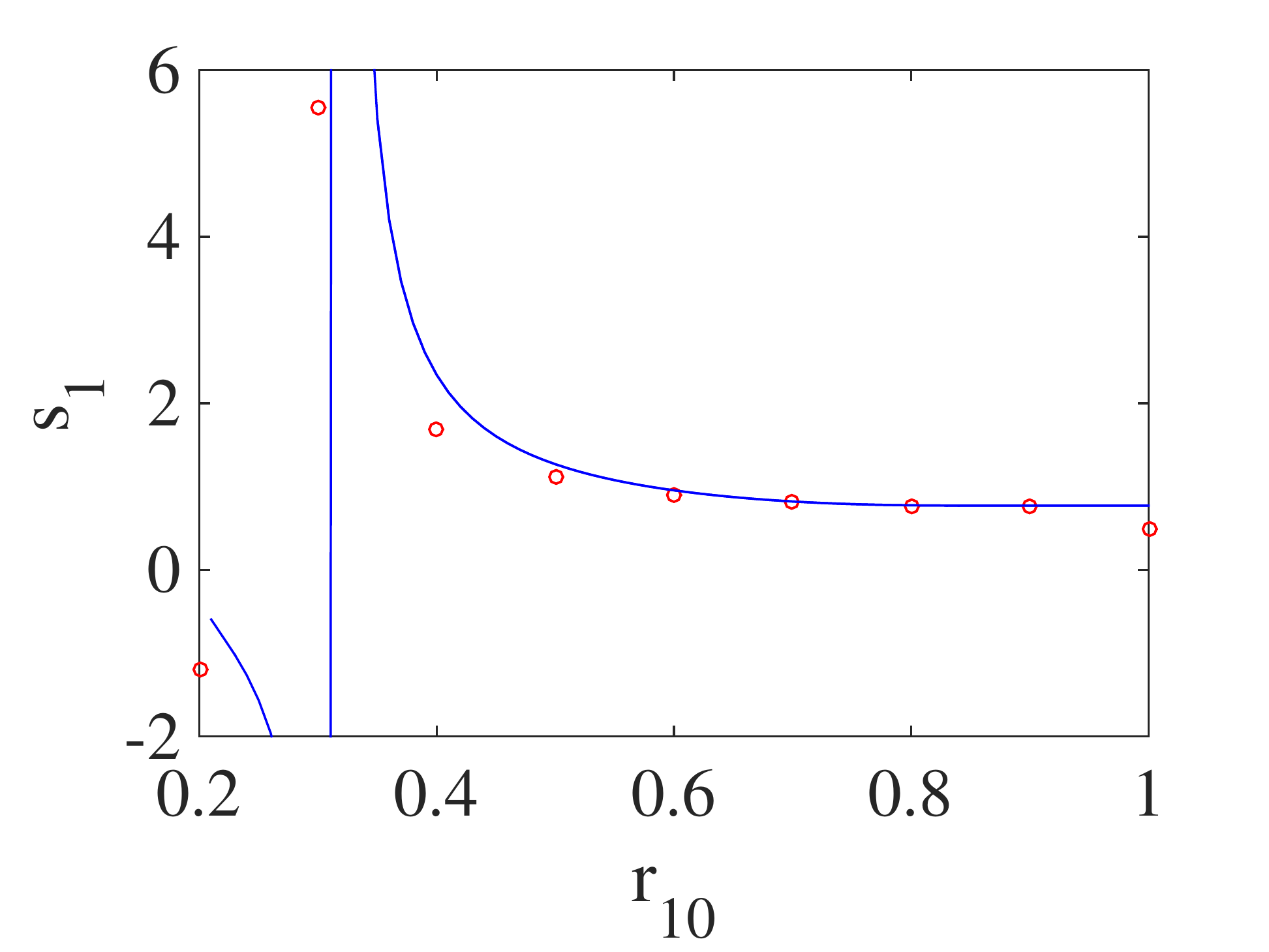} &   
    \includegraphics[valign=m,width=.3\textwidth]{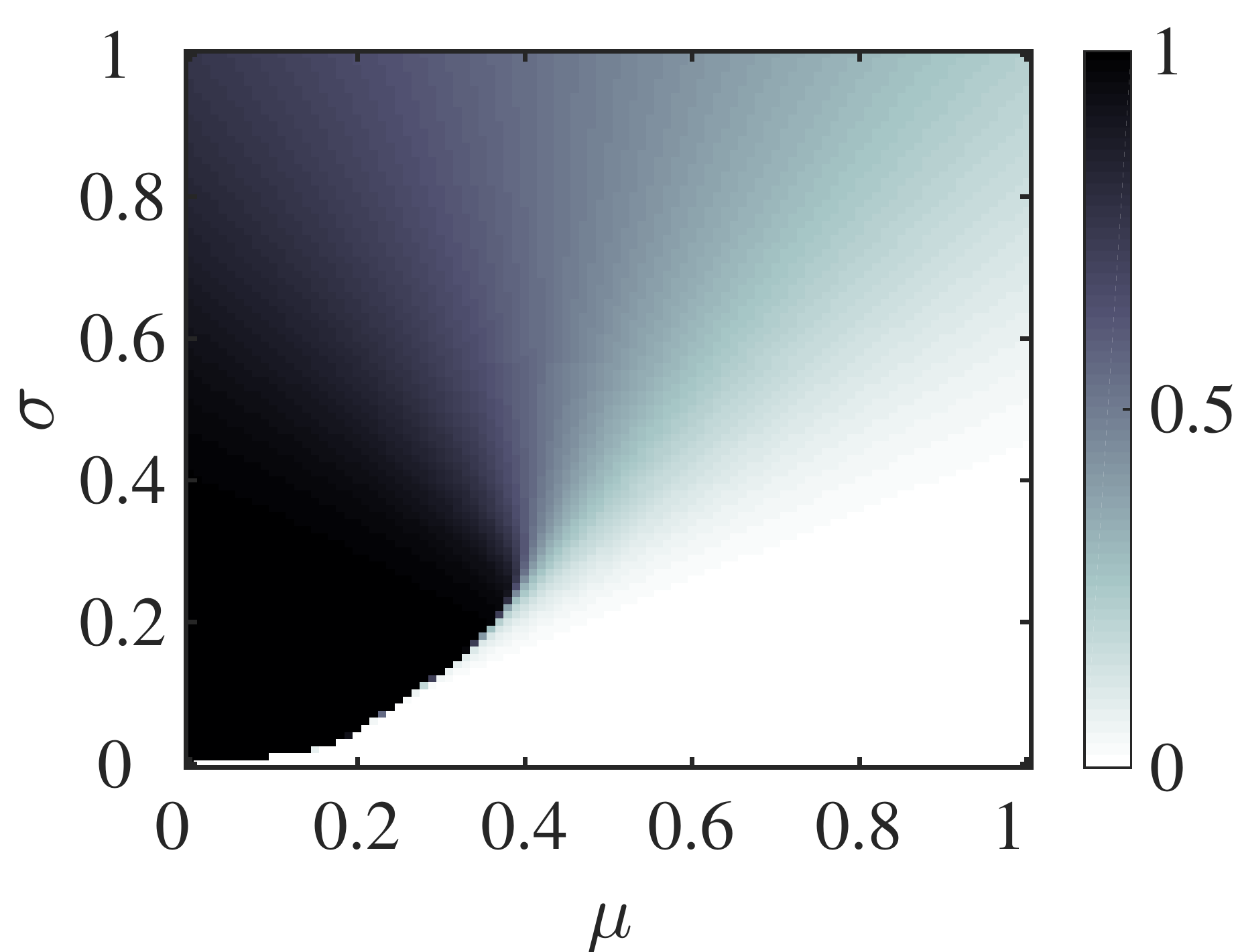} &
    \includegraphics[valign=m,width=.32\textwidth]{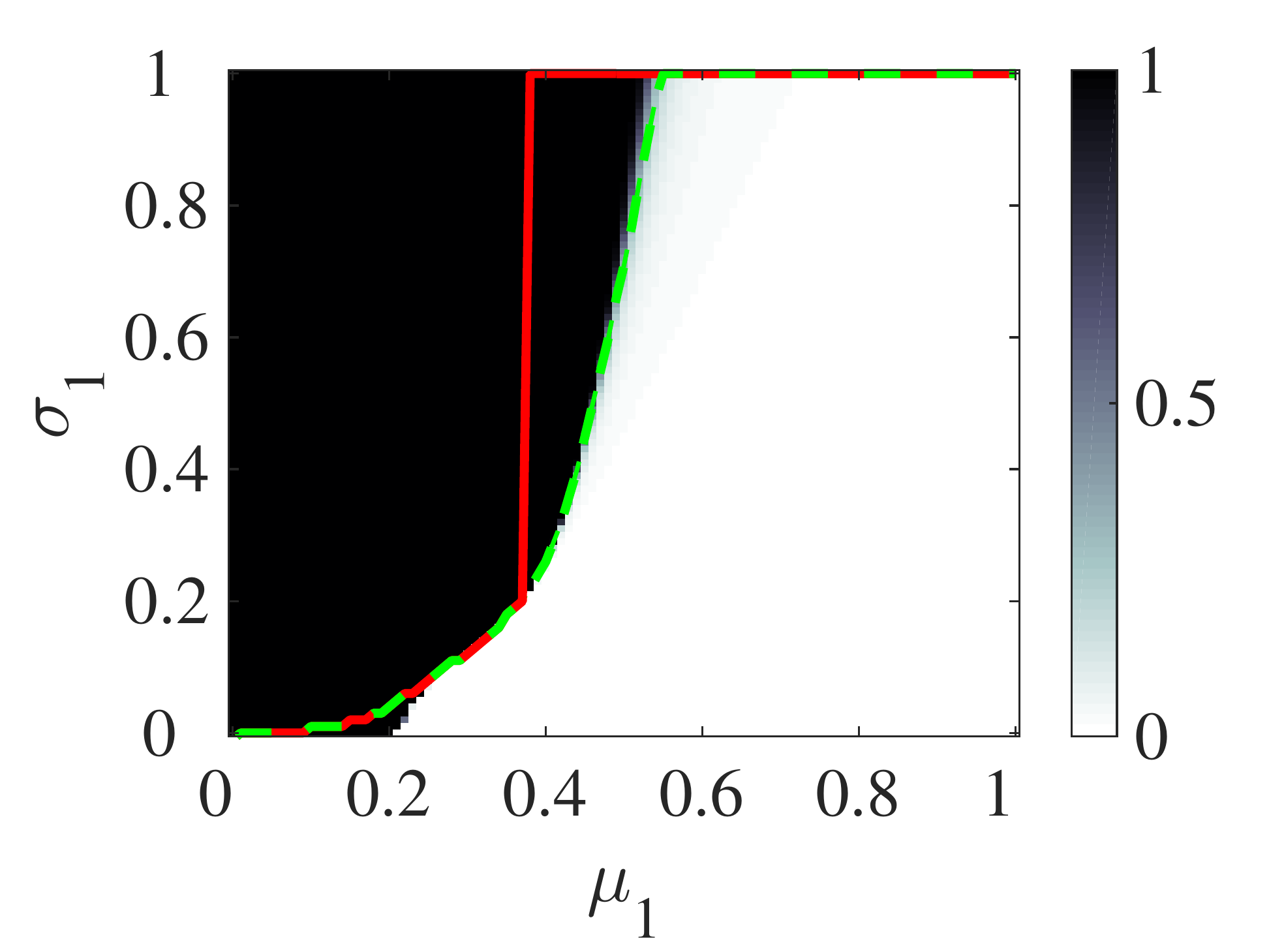}\\
    \end{tabular}
  \caption{\textit{\emph{(Left):} Red Circles: Slope $s_1$ of the linear phase transition obtained by linear regression of $\rho^*_{0}$ that we calculated by solving the fixed point equations (\ref{eq:fixp}) numerically for different values of $r_{10}$. The blue curve shows the approximate values for the slopes provided by Eqn. \ref{eq:initSlope}.}
  \textit{\emph{(Middle):} Phase diagram for the final cascade size on a single layer without threshold feedback for Poisson random graphs with average degree $z = 5$ and normally distributed thresholds $F_{\theta} \sim \mathcal{N}\left(\mu, \sigma^2\right)$.}
  \textit{\emph{(Right):} Final cascade size on layer 0, i.e. $\rho^*_{0}$, for $r_{10} = 0.3$. The red line marks the phase transition defined by the criterion $\rho^*_{\text{single}} \geq \rho^{c}_{s,1}$. The green line corresponds to $\rho^*_{\text{single}} \geq 0.45$.}
  }
   \label{fig:slope}
\end{figure}

In Fig. \ref{fig:z1_5_z2_5_muc_03_sigc_01_agg} we observe that the sharp phase transition scales almost linearly for large enough coupling strength $r_{10}$, i.e.
\begin{equation}
\sigma_1 =  s_1 \mu_1 + s_0 \quad \mathrm{for}\; \mu_1 \geq 0.4, 
\; r_{10}\geq r_{10}^{c} \label{eq:5}
\end{equation}
In order to determine the scaling parameter $s_{1}$, we take the phase transition line from the phase diagrams of Fig. \ref{fig:z1_5_z2_5_muc_03_sigc_01_agg}, which were also calculated for $r_{10}=$0.3, 0.5, 0.6, 0.7, 0.9. 
These lines are approximated by a linear regression, and the slopes obtained this way are plotted in Fig. \ref{fig:slope} (left, red dots) against the coupling strength $r_{10}$. 
We observe a non-monotonous dependence of  the slope, with a saturation effect for large $r_{10}$. 
Thus, we want to determine $r_{10}^{c}$ and $s_{1}(r_{10})$ for $r_{10}\geq r_{10}^{c}$ . 

To better understand this dependence, we perform a linear approximation of $L_0$ in Eqn. (\ref{eq:fixp}) as
\begin{equation}
L_0\left(\pi_0\right) \approx a + b\pi_0
\label{eq:4}
\end{equation}
which is valid for small values of the failure probability $\pi_0$. 
Instead of solving the full system of fixed point equations Eqn. \eqref{eq:fixp}, we only solve the linearized system in order to deduce criteria for the growth behavior of $\pi_{0}$ in the fixed point  iterations.  
The fixed point of the linear equation \eqref{eq:4} exceeds one if 
\begin{align}\label{eq:linFix}
\frac{a}{1-b} \geq 1
\end{align}
Under this condition, initial failures in both layers result in large cascades in layer 0. 

Eqn. \eqref{eq:linFix} leads to conditions for the parameters $\mu_{1}^{c}$, $\sigma_{1}^{c}$ and $r_{10}^{c}$ if we know the expressions for $a$ and $b$. 
These are obtained from the linearization of Eqn. \eqref{eq:fixp}:
\begin{align}\label{eq:linParam}
a = & \mathbb{P}\left(s_0 = 1 \vert k_0, n_0=0, \rho^*_{s,1}\right),  \nonumber \\
b = & \sum^{\infty}_{k_0=1} \frac{k_0 (k_0 - 1)}{z_0} p_0\left(k_0\right) \times  \\ 
& \times \left[\mathbb{P}\left(s_0 = 1 \vert k_0, n_0=1, \rho^*_{s,1}\right)- \mathbb{P}\left(s_0 = 1 \vert k_0, n_0=0, \rho^*_{s,1}\right)\right].
\nonumber
 \end{align}
These parameters depend on the failure probability $\rho_{s,1}^{*}$ of a node in the other layer, i.e. layer 1. 

For the further discussion, we first concentrate on the worst case scenario  $\rho_{s,1}^{*}=1$, i.e. all nodes failed on layer 1. 
We are interested in identifying the regime where even in the worst case, cascades in layer 1 do not propagate to layer 0.
In other words, we want to approximate the critical coupling $r^{c}_{10}$ 
above which we can observe regions of high systemic risk in the left column of Fig. \ref{fig:mu2_r21_plot_mu1_03_sigma1_01_sigma2_03}. 
With  $\rho_{s,1}^{*}=1$ the parameters $a$ and $b$ simplify to
\begin{align}
\begin{split}
& a = F_0(0), \\
&  b = \sum^{\infty}_{k_0=1} \frac{k_0 (k_0 - 1)}{z_0} p_0\left(k_0\right) \left[ F_0\left( \frac{1}{k_0 \left(1-r_{10}\right)}\right) - F_0(0) \right].
\end{split}
\end{align}
We note that these are independent of $\mu_{1}$, $\sigma_{1}$.
For the set of parameters used in Fig. \ref{fig:mu2_r21_plot_mu1_03_sigma1_01_sigma2_03}, we obtain with the help of Eqn. \eqref{eq:linParam} and Eqn. (\ref{eq:linFix}) the value of the critical coupling $r^{c}_{10} = 0.204$. 
In comparison with the numerical calculation $r_{10} = 0.18$, which takes the full set of fixed point equations into account, this is a good approximation. 
That means large cascades can propagate from layer 1 to layer 0 above a critical coupling strength  $r^{c}_{10} \simeq 0.2$.
This holds independently of $\mu_{1}$ and $\sigma_{1}$.

Next we explain the dependence $s_{1}(r_{10})$ as shown in Fig. \ref{fig:z1_5_z2_5_muc_03_sigc_01_agg}. By this we estimate the slope of the phase transition line (which is of the form $\sigma_1 \propto   s_1 \mu_1$). Again, we deduce this relation from Eqs. \eqref{eq:linFix}, \eqref{eq:linParam}, but this time we cannot use  
$\rho^*_{s,1}=1$. 
For $\rho^*_{s,1}\neq 1$, we automatically obtain a dependence on the threshold parameters $\mu_1$ and $\sigma_1$ of layer 1. 
Combining Eqn. (\ref{eq:linParam}) and Eqn. (\ref{eq:linFix}), we can define a critical value $\rho^{c}_{s,1}$ such that Eqn. (\ref{eq:linFix}) is satisfied for $\rho^*_{s,1} \geq \rho^{c}_{s,1}$:
\begin{align}
 \rho^{c}_{s,1} := \frac{1 + F_0(0) (z_0 -1) - c_0(0)}{c_0(r) - c_0(0)},
\end{align}
where 
\begin{align}
 c_0(r) := \sum^{\infty}_{k_0=1} \frac{k_0 (k_0 - 1)}{z_0} p_0\left(k_0\right)F_0\left( \frac{1}{k_0 \left(1-r_{10}\right)}\right).
\end{align}
This is a reformulation of Eqn. \eqref{eq:linFix}, which gives us an estimation of when to expect large cascades in layer 0.
Still, we do not know  $\rho^*_{s,1}$ without solving the full system of fixed point equations \eqref{eq:fixp}.
In the following we test two proxies for $\rho^*_{s,1}$  which lead us to a linear dependence between $\mu_1$ and $\sigma_1$.

(1) The initial value $\rho_{s,1} (t=0) = F_1(0) = \Phi\left(-{\mu_1}/{\sigma_1} \right)$ is already enough to determine the growth of $\pi_0$ in the early stages of the fixed point iterations. 
Especially for larger values of $r_{10}$, a large cascade in layer 0 might be already triggered just by the initial failures in layer 1. 
Therefore, we set $\rho^{c}_{s,1} \simeq F_1(0)$, which results in
\begin{align}\label{eq:initSlope}
 \sigma_1 \propto - \mu_1 / \Phi^{-1}\left( \rho^{c}_{s,1} \right),
\end{align}
i.e., 
\begin{equation}
  \label{eq:6}
  s_{1} =  - {1}/{\Phi^{-1}\left( \rho^{c}_{s,1} \right)}
\end{equation}
where $\Phi^{-1}$ denotes the cumulative distribution function of the standard normal distribution. 
We plot this slope using a blue line in Fig. \ref{fig:slope}(left), to demonstrate the good agreement of our approximation with the numerical slopes $s_{1}$ for $r_{10} \geq 0.4$. 

(2) For smaller coupling strengths $r_{10}$ we test a second proxy for $\rho^*_{s,1}$. 
A lower bound for $\rho^*_{s,1}$ is given by the final cascade size on a single layer $\rho^*_{\text{single}}$, where no feedback mechanism with another layer exists. 
That means, the other layer cannot further amplify the failures. 
Cascades in single layers have been studied in~\cite{Burkholz2015}. 
We use their approach to plot a phase diagram for the final cascade size for Poisson random graphs with average degree $z = 5$, shown in the middle panel of Fig. \ref{fig:slope}.

When we compare the plots of Fig. \ref{fig:z1_5_z2_5_muc_03_sigc_01_agg} (left) with Fig. \ref{fig:slope} (middle), we observe that in the parameter regions in Fig. \ref{fig:z1_5_z2_5_muc_03_sigc_01_agg} (left) where no cascades occur, also no cascades occur in Fig. \ref{fig:slope} (middle). 
This holds only for strong couplings $r_{10}$. 
Consequently, in this limit, our second proxy for $\rho^*_{s,1}$ gives similar results as our first proxy. 

However, the second proxy for $\rho^*_{s,1}$, is the more general one as we can also correctly cover the transition line for values of $0.2 \leq \mu_{1}\leq 0.4$. 
In this range, the transition line is largely independent of the value $r_{10}$. 
It basically coincides with the transition line for $\rho^*_{\text{single}}$, the fraction of failed nodes for single layer networks. 

For values of $\mu_{1}$ larger than 0.4 and for weak coupling, e.g. for $r_{10} = 0.3$, we observe in  Fig. \ref{fig:slope} (right, red line) that the transition line is no longer accurately described. 
The condition $\rho^*_{\text{single}} \geq \rho^{c}_{s,1}=0.55$ gives only an \emph{upper bound} for the transition line, but not a lower one. 
We observe that already cascade sizes of $\rho^*_{\text{single}} \geq 0.45$ lead to large cascade amplification on layer 0, due to nonlinear effects.
I.e. the green dotted line in  Fig. \ref{fig:slope} (right) gives an almost perfect match with the numerically calculated transition line. 

That means, we can even generalize our statement that  the transition line for $\rho^*_{\text{single}}$ determines the phase transition. 
This holds not only for small $\mu_{1}$, but also for larger $\mu_{1}$ and a broad range of the coupling strength $r_{10}$. 

\section{Conclusions}
\label{sec:Conclusion}

Our paper essentially addresses the problem of whether systemic risk is increased or decreased if, instead of a single-layer network, a two-layer representation is used. 
In the latter, the nodes of the network appear on two layers 0 and 1 with different properties. 
Specifically, they have different degree distributions $p_0\left(k_{0}\right)$, $p_1\left(k_{1}\right)$ and different threshold distributions $F_0\left(\theta_{0}\right)$, $F_1\left(\theta_{1}\right)$. 
There is an asymmetric coupling between the two layers such that nodes that failed on layer 0 also fail on layer 1. 
However, nodes that did not fail on layer 0, may still fail on layer 1 because of a cascade dynamics on layer 1. 
In this case, their failing threshold on layer 0 is reduced by a fraction $r_{10}$, where $r_{10}$ denotes the coupling strength between the two layers. 

The mutual feedback between the two layers can then result in the amplification of failure cascades, which we study analytically. 
We can calculate a variable $\rho_{l}^{*}$ which is the final fraction of failed nodes on each layer $l\in \{0,1\}$. 
Our measure of systemic risk is $\rho^{*}\equiv \rho_{0}^{*}$, i.e. we only consider whether nodes have failed on layer 0.
Obviously, if $r_{10}$ is small, no failure cascade in layer 1 can propagate to layer 0.
In this case, whether or not we observe failure cascades only depends on the conditions in layer 0. 
These conditions are expressed by the parameters of the threshold distribution, $\mu_{0}$ and $\sigma_{0}$, which are chosen so that no failure cascade occurs. 
By varying $r_{10}$, we then study the impact of failures on layer 1 on failures on layer 0. 
We derive an analytical approach  to calculate $\rho^{*}_l$, which leads to a system of coupled fixed point equations solved numerically. 

Our results are visualized by means of phase diagrams that show, for various parameter constellations, the value of the main risk measure $\rho^{*} = \rho^{*}_0$. 
The most prominent feature is the existence of a very sharp phase transition between a regime of \emph{no} systemic risk and a regime of full collapse.
This means that small changes in the parameters $\mu_{1}$ and $\sigma_{1}$, describing the threshold distribution on layer 1, can lead to an abrupt regime shift. 
Our task was then to approximate this line of transition in terms of the coupling strength $r_{10}$ and the parameters of the threshold distribution. 
Subsequently, we use these insights to compare the systemic risk in single-layer and two-layer networks. 

The derivation of mathematical approximations for the phase transitions that are compared, and confirmed, by numerical solutions of the full problem has a value on its own. 
Here, we focus on the conclusions that can be made based on these calculations. 
First of all, we understand that systemic risk is reduced in the two-layer network only if the coupling strength between the two layers is rather small. 
Above a critical value $r_{10}^{c}=0.2$, which is also estimated analytically, failures on layer 1 are amplified on layer 0 and thus lead to an increase of the systemic risk. 
If we compare this with the reference case of a single layer network, we find that the systemic risk is smaller there, for most ranges of the parameters. 
Hence, above a critical coupling, it is not beneficial to split the network into two layers, if systemic risk shall be mitigated. 

Our findings can be applied to a scenario where firms have to decide whether to split their business into a less risky core business, essential for their survival, and a subsidiary business, which can be more risky. 
Such a split leads to the representation of the (same) firm on two different levels 0 and 1, where the latter has a larger risk of failure cascades. 
This decision only leads to less failure risk for the firm if the coupling between the core business layer and the subsidiary business layer is weak enough. 
Close to the transition line between the no-risk and the high-risk regime, slight changes in the firms' failure thresholds or in the coupling between the different businesses may potentially cause the system to collapse completely. 

There is another important conclusion to be drawn for firms that already have their activities split between a less risky core business and a more risky subsidiary business. 
If such firms estimate systemic risk based on an aggregated network representation with only one layer instead of the two-layer representation, they may systematically underestimate the real risk, in particular if the coupling is still strong. 
As we have shown, under such conditions the systemic risk in the aggregated network is lower than in the two-layer network. 
Hence, there can be drastic consequences from drawing conclusions based on an inappropriate aggregated picture. 

\paragraph*{Acknowledgements.}
R. Burkholz acknowledges support by the ETH48 project, A. Garas and F. Schweitzer acknowledge support by the EU-FET project MULTIPLEX 317532. The authors thank I. Scholtes for discussions.

\bibliographystyle{sg-bibstyle}
\bibliography{LBGS-biblio}

\end{document}